\documentclass[3p,a4paper,11pt,review]{elsarticle}

\usepackage{lineno}
\usepackage{hyperref}
\usepackage{latexsym}
\usepackage{color}
\usepackage[table]{xcolor}
\usepackage{mathrsfs}
\usepackage{amssymb,amstext,amsfonts,amsmath,amsthm}
\usepackage{bm}
\usepackage{subcaption}
\usepackage[utf8]{inputenc} 
\usepackage[english]{babel}
\usepackage{hyphenat}
\usepackage{calc}
\usepackage{graphicx,import}

\usepackage{soul}

\usepackage{caption}
\usepackage{subcaption}

\usepackage{arydshln}
\usepackage{natbib}
\usepackage[misc,geometry]{ifsym} 
\usepackage{empheq}
\usepackage{longtable}
\usepackage{comment}

\biboptions{authoryear}

\usepackage{tikz}

\def\bbR{{\mathbb R}}

\def\tr[#1]{\textcolor{red}{#1}}
\def\tg[#1]{\textcolor{green}{#1}}
\def\tb[#1]{\textcolor{blue}{#1}}

\newcommand{\bfzero}{{\bf 0}}

\newcommand{\bfc}{{\bf c}}
\newcommand{\bfC}{{\bf C}}
\newcommand{\bfd}{{\bf d}}

\newcommand{\bfe}{{\bf e}}
\newcommand{\bfE}{{\bf E}}
\newcommand{\bff}{{\bf f}}

\newcommand{\bfG}{{\bf G}}

\newcommand{\bfI}{{\bf I}}

\newcommand{\bfH}{{\bf H}}

\newcommand{\bfL}{{\bf L}}
\newcommand{\bfm}{{\bf m}}

\newcommand{\bfN}{{\bf N}}
\newcommand{\bfp}{{\bf p}}
\newcommand{\bfP}{{\bf P}}

\newcommand{\bfr}{{\bf r}}
\newcommand{\bfR}{{\bf R}}
\newcommand{\bfRzero}{{\bf R_0}}

\newcommand{\bft}{{\bf t}}

\newcommand{\bfu}{{\bf u}}

\newcommand{\bfv}{{\bf v}}

\newcommand{\bfx}{{\bf x}}

\newcommand{\bfw}{{\bf w}}

\newcommand{\bsthetaloc}{\boldsymbol{ \overline{ \theta } }}

\def\x{{\mathbf x}}

\renewcommand{\leq}{\leqslant}

\usepackage{lineno}
\newcommand*\patchAmsMathEnvironmentForLineno[1]{%
	\expandafter\let\csname old#1\expandafter\endcsname\csname #1\endcsname
	\expandafter\let\csname oldend#1\expandafter\endcsname\csname end#1\endcsname
	\renewenvironment{#1}%
	{\linenomath\csname old#1\endcsname}%
	{\csname oldend#1\endcsname\endlinenomath}}%
\newcommand*\patchBothAmsMathEnvironmentsForLineno[1]{%
	\patchAmsMathEnvironmentForLineno{#1}%
	\patchAmsMathEnvironmentForLineno{#1*}}%
\AtBeginDocument{%
	\patchBothAmsMathEnvironmentsForLineno{equation}%
	\patchBothAmsMathEnvironmentsForLineno{align}%
	\patchBothAmsMathEnvironmentsForLineno{flalign}%
	\patchBothAmsMathEnvironmentsForLineno{alignat}%
	\patchBothAmsMathEnvironmentsForLineno{gather}%
	\patchBothAmsMathEnvironmentsForLineno{multline}%
}

\usepackage[]{algorithm2e}

\usepackage{tikz}
\usetikzlibrary{shapes,arrows}

\usepackage{relsize}

\definecolor{darkcandyapplered}{rgb}{0.64, 0.0, 0.0}
\definecolor{darkred}{rgb}{0.55, 0.0, 0.0}

\def\tr[#1]{\textcolor{darkcandyapplered}{#1}}
\def\tb[#1]{\textcolor{blue}{#1}}

\journal{ }

\usepackage{bbm}

\usepackage{enumitem,amssymb}
\newlist{todolist}{itemize}{2}
\setlist[todolist]{label=$\square$}
\usepackage{pifont}

\begin{document}


\begin{frontmatter}
	
  \title{A consistent co-rotational formulation for aerodynamic nonlinear analysis of flexible frame structures}

\author[1]{Mauricio C. Vanzulli}
\author[2]{Jorge M. Pérez Zerpa}

\address[1]{Instituto de Ingeniería Mecánica y Producción Industrial, Facultad de Ingeniería, Universidad de la República, Montevideo, Uruguay}
  
\address[2]{Instituto de Estructuras y Transporte, Facultad de Ingeniería, Universidad de la República, Montevideo, Uruguay}  

\begin{abstract}
	
The design of structures submitted to aerodynamic loads usually requires the development of specific computational models considering fluid-structure interactions. Models using structural frame elements are developed in several relevant applications such as the design of advanced aircraft wings, wind turbine blades or power transmission lines. In the case of flexible frame structures submitted to fluid flows, the consistent computation of inertial and aerodynamic forces for large displacements is a challenging task. In this article we present a novel formulation for the accurate computation of aerodynamic forces for large displacements and rotations using the co-rotational approach, the \emph{quasi-steady theory} and the principle of virtual work. This formulation is coupled with a reference consistent co-rotational formulation for computing internal and inertial forces, providing a unified set of nonlinear balance equations. A numerical resolution procedure is proposed and implemented within the open-source library ONSAS. The proposed formulation and its implementation are validated through the resolution of five examples, including a realistic wind turbine analysis problem. The results provided by the proposed formulation are compared with analytic solutions and solutions obtained using a lumped mass approach. The proposed formulation provides accurate solutions for challenging numerical problems with large displacements and rotations.

\end{abstract}

\begin{keyword}
	Co-rotational formulation \sep Nonlinear dynamics \sep Quasi-steady theory
 \sep Finite Element method \sep  Fluid-Structure Interaction \sep Open-source software
  \end{keyword}	
\end{frontmatter}

\section{Introduction}

Nonlinear structural dynamics problems are formulated in a vast and diverse set of applications such as: developing new offshore wind turbines \citep{Ahsan2022a}, designing suspended bridges or plane wings \citep{Zhao2018,Binder2021}, predicting failures in power transmission lines \citep{Solari2020}, reducing fruit production losses \citep{Cataldo2013} or even studying the movement of aquatic plants \citep{Gosselin2019}. In all of these applications, structures can be modeled using frame elements, and are also submitted to loads caused by the interaction with fluid flows. The development of an efficient and accurate numerical method for the resolution of this type of problems, and its open-source implementation, are the main motivations of this article.

Structural design standards have a limited range of application, and are not applicable to most of the  problems mentioned above \citep{Duranona2019,Salehinejad2021}. Given this limitation, alternative approaches are mainly based on experimental tests \citep{Cataldo2013} or numerical simulations \citep{Stengel2017b}.  Experimental tests might be expensive and/or challenging to design, therefore, new numerical methods for accurate structural dynamics simulations are actively developed \citep{Forets2022}.

The Finite Element Method (FEM) \citep{Zienkiewicz1972} has become the gold-standard for computational modeling in structural analysis  in numerous disciplines. For frame structures, the co-rotational approach has shown several advantages, including a more versatile and less intricate mathematical formulation \citep{Crisfield1997}. This approach is basically based on splitting the element deformation in two: one rigid movement and one local deformation \citep{Belytschko1979}. Different co-rotational formulations were gradually developed for its use in static analysis \citep{Nour-Omid1991}, considering instability \citep{Battini2002} or dynamic analysis \citep{Le2011a}. Lately, a consistent formulation for three dimensional nonlinear dynamic analysis of frame structures was presented \citep{Le2014}, enabling to accurately model movements with large displacements and rotations using a reduced number of elements. In \citep{wang2020high} it is shown that, for structures submitted to large rotations, the consistent formulation is considerably more accurate and efficient than the lumped mass approach.

In the last decades different frame analysis formulations were used for the mentioned applications of interest. In \citep{Desai1995}, a 3D nonlinear three-node isoparametric element is used for modeling overhead transmission lines movement, considering a consistent mass matrix for linear inertial terms. With the same purpose, in \citep{Shehata2005a} a three dimensional linear frame element was used to simulate cable elements. In \citep{Maalawi2002}, a formulation considering nonlinear internal forces with a lumped mass matrix for linear inertial terms was used for modeling wind turbine blades. In \citep{Gosselin2010} and \citep{Hassani2016} differential equations modeling fluid flows and drag forces over frame structures were integrated with nonlinear beam formulations using Euler-Bernoulli and Kirchhoff theories, respectively. In \citep{Piccardo2016} the \emph{quasi-steady theory} and an equivalent beam model with a lumped mass matrix was employed to analyze galloping effects on buildings. Using the same aerodynamic theory, discrepancies between linear formulations and experimental measurements in transmission lines were addressed in \citep{Foti2018}, considering a co-rotational framework with a lumped mass matrix. Regarding nonlinear geometric analysis of wind turbine blades, the linearized equations of motion were solved in \citep{FaccioJunior2019}, obtaining a good level of agreement between the \emph{exact beam theory} and formulations using shell elements. In \citep{DeBreuker2011a,Macquart2020}, static co-rotational formulations were used to simulate morphing or highly flexible wings, highlighting the computational efficiency to validate experimental results.

Regarding the availability of software for the numerical resolution of these problems, three specific tools can be mentioned: RIFLEX, FAST and HAWC2. RIFLEX is a proprietary software developed for fluid-structure interaction problems. It uses a co-rotational approach for modeling frame elements, a linear consistent mass matrix, a Rayleigh damping matrix and allows to compute mass, shear and elastic centers \citep{Cheng2017,delhaye2017effect}. FAST is a modular open-source framework for fluid structure numerical simulations. This software uses beam elements based on the \emph{exact beam theory}, implemented with a consistent Timoshenko mass matrix \citep{Wang2017}. HAWC2 is a proprietary software for aeroelastic simulation of wind turbines, developed using linear anisotropic Timoshenko beam elements \citep{Kim2013}.

In this work we present a unified formulation for consistent co-rotational analysis of frame structures submitted to aero-dynamic forces. The effect of the fluid interaction is included by considering the \emph{quasi-steady theory}, co-rotational kinematics and the principle of virtual work. Moreover, we implement the formulation as part of the open-source structural analysis tool ONSAS\footnote{\href{http://www.onsas.org}{www.onsas.org}} \citep{ONSAS}. We perform numerical analyses for different flow conditions, cross-sections and magnitudes of displacements and rotations, studying changes in mesh sizes and number of Gauss numerical integration points. The formulation and its implementation are validated through the resolution of five numerical examples, including a realistic wind turbine problem. All the scripts used in the numerical examples are available, allowing any user to reproduce the results or inspect, analyze or modify the models.

This article is organized as follows. In Section~\ref{sec:prelimiaries} the basic concepts of the co-rotational frame formulation and the main hypotheses of the \emph{quasi-steady theory} are described. In Section~\ref{sec:methodology}, the proposed formulation is presented, with a corresponding numerical procedure for the resolution of the balance equations. In Section~\ref{sec:numericalResults} the numerical results obtained for five validation problems are presented, and in Section~\ref{sec:conclusions}, the conclusions obtained are presented.

\section{Preliminaries}\label{sec:prelimiaries}

In this section, the fundamental concepts of the co-rotational frame analysis approach are described. The main kinematic identities and the nonlinear dynamics internal and inertial forces are briefly presented considering \citep{Battini2002,Le2014} as references, respectively.

\subsection{Co-rotational kinematics} \label{subsec:PreCorotKineamitcs}

The main concepts behind the co-rotational approach are the use of different systems of coordinates and the application of the principle of virtual work. Given a two-node frame element and two systems of coordinates (global and local), a vector of generalized nodal displacements $\bfd$ can be represented in the global system of coordinates as $\bfd_g$, and in the local system as $\bfd_\ell$. A vector of nodal forces $\bff$ (which might depend on the displacements), can also be written in those corresponding systems, and the principle of virtual work can be stated as:
\begin{equation}\label{eqn:generalPVW}
(\delta \bfd_\ell)^T\bff_{\ell} = (\delta \bfd_g)^T\bff_{g},
\end{equation}
for any vector of virtual displacements $\delta \bfd$.

\begin{figure}[htb]	
	\centering
	\def\svgwidth{0.65\textwidth}
	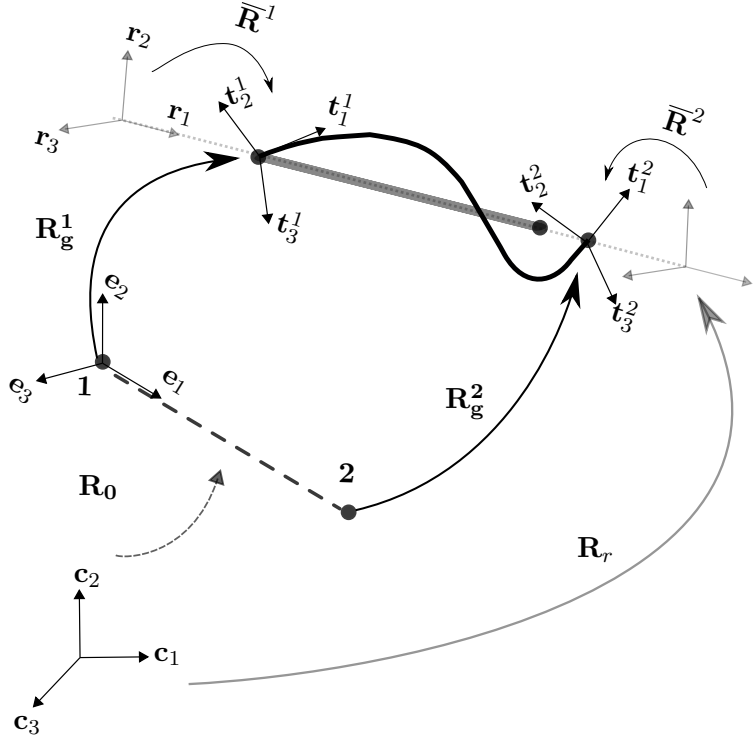
	\caption{Diagram of the co-rotational framework: reference configuration (dashed line), rigid-rotation configuration (gray solid line) and total-deformed configuration (black solid line).}
	\label{fig:preIlusCorrot}
\end{figure}

In the co-rotational approach three configurations are defined as shown in Figure~\ref{fig:preIlusCorrot}: a reference configuration (without deformation), a rigid-rotation configuration and the total-deformation configuration. As it is shown, four systems of coordinates are defined: $\left\{ \bfc_i \right\}$,  $\left\{\bfe_i\right\}$,  $\left\{\bfr_i\right\}$ and  $\left\{\bft_i\right\}$, corresponding to the canonical, reference, rigid-rotation and total-deformed configurations, respectively. Orthogonal matrices $\bfR_0$, $\bfR_g$, $\bfR_r$ and $\overline{\bfR}$ can also be defined as shown in Figure~\ref{fig:preIlusCorrot}, to rotate the base vectors of these systems of coordinates.

\begin{figure}[htb]
	\centering
	\def\svgwidth{0.65\textwidth}
	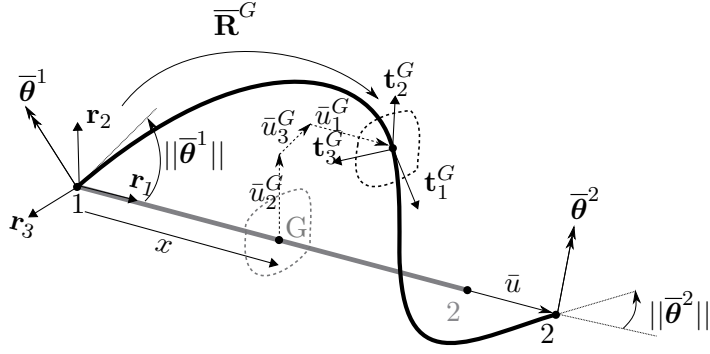
	\caption{Local displacements from rigid-rotation to total-deformed configuration.}
	\label{fig:preIlusLocalDisp}
\end{figure}

The column vector of nodal displacements written in the canonical system $\left\{ \bfc_i \right\}$ is denoted as $\bfd_g = [(\bfu^1)^T, (\bfw^1)^T, (\bfu^2)^T,   (\bfw^2)^T]^T$, where $\bfu^i$ and $\bfw^i$ are the column vectors of linear displacements and rotations, respectively, of node $i$. In the co-rotational approach, the displacements of the element are also written considering the system of coordinates $\left\{ \bfr_i \right\}$, where the local extension and the nodal rotations are grouped as $\bfd_\ell=[ \overline{u}, (\bsthetaloc^1) ^T , (\bsthetaloc^2) ^T ] ^T$. The extension is given by $\overline{u} = l_n - l_0$ where $l_n$ and $l_0$ are the deformed and reference lengths of the element, and the local rotations are given by the vectors $\bsthetaloc^i$ as it is shown in Figure~\ref{fig:preIlusLocalDisp}.

\setcounter{MaxMatrixCols}{20}

In order to apply the Principle of Virtual Work, the variations of the generalized displacements (given by the virtual displacements) in different systems of coordinates need to be computed. The variation of the local extension can be written as:
\begin{equation}\label{eq:preDefR}
\delta \overline{u} = \bfr~ \delta \bfd_g,  \qquad\bfr = [ -\bfr_1^T~ \bfzero_{1\times 3}~ \bfr_1^T~ \bfzero_{1 \times 3}  ],
\end{equation}
and for the vectors of rotations:
\begin{equation}\label{eq:PreLocalAngVar}
\begin{bmatrix}
\delta\bsthetaloc^1\\
\delta\bsthetaloc^2
\end{bmatrix}
=\bfP\bfE^T \delta \bfd_g,
\qquad
\bfP =  \begin{bmatrix}
\bfzero_{3\times 3} &\bfI  & \bfzero_{3\times 3} &\bfzero_{3\times 3} \\
\bfzero_{3\times 3} & \bfzero_{3\times 3}  &\bfzero_{3\times 3}  & \bfI
\end{bmatrix}-\begin{bmatrix}
\bfG\\
\bfG
\end{bmatrix},
\end{equation}
where $\bfzero_{i\times j}$ represents a matrix of zeros with $i$ rows and $j$ columns (the sub-indexes are omitted for the $3\times3$ case), $\bfE$ is a matrix given by: 
\begin{equation}\label{eq:preDefE}
\bf{E}=\begin{bmatrix}
\bf{R_r}& \bf{0}   & \bf{0}   & \bf{0} \\
\bf{0}  & \bf{R_r} & \bf{0}   & \bf{0}\\
\bf{0}  & \bf{0}   & \bf{R_r} & \bf{0} \\
\bf{0}  & \bf{0}   & \bf{0}   & \bf{R_r}
\end{bmatrix},
\end{equation}
and $\bfG$ is a matrix given by:
\begin{equation}\label{eq:preDefG}
\bf{G} =
\begin{bmatrix}
0 &  0      &  \frac{p_1}{p_2 l_n} &  \frac{p_{12}}{2p_2} &-\frac{p_{11}}{2p_2}  &  0  & 
0  &  0    & -1/l_n & 0 & 0 & 0 \\
0 &  0      &   1/l_n   &       0       &      0       &  0   & %
0  &  0     &-\frac{p_1}{p_2 l_n}  & \frac{p_{2}}{2p_2} & -\frac{p_{21}}{2p_2} &    0 \\
0 & -1/l_n  &      0    &       0       &      0       &  0 & %
0  &  1/l_n &       0   &      0      &     0        &    0
\end{bmatrix},
\end{equation}
with $p_{ij}$ being the $j$-th entry of the vector $\bfp_i$ defined by:
\begin{equation}
	\bfp_i=\bfR_g^i\bfR_0[0,1,0]^T \quad i=1,2,
\end{equation}
and $p_j$ being the $j$-th entry of the vector $\bfp$ defined by $\bfp=\frac{1}{2}(\bfp_1+\bfp_2)$.

For a cross-section located at the position $x$, as shown in Figure~\ref{fig:preIlusLocalDisp}, with centroid $G$ and deformed base $\left\{\bft_i^G\right\}$, the variations of the displacements and rotations can also be written in local and global systems, as:
\begin{equation} \label{eq:preCompactTransverseDisp}
\begin{bmatrix}
0     		\\
\bar{u}_2^G \\
\bar{u}_3^G
\end{bmatrix}
=\bf{P}_1
\begin{bmatrix}
\bsthetaloc^1\\
\bsthetaloc^2
\end{bmatrix},
\qquad
\bf{P}_1 =
\begin{bmatrix}
0  &   0    & 0 	& 0  	& 	0 	& 0   \\
0  &   0    & N_3 	& 0 	& 	0 	& N_4 \\
0  &  -N_3  & 0 	& 0 	& -N_4 	& 0
\end{bmatrix},
\end{equation}
and
\begin{equation}\label{eq:preCompactTransverseAngles}
\begin{bmatrix}
\bar{\theta}_1^G \\
\bar{\theta}_2^G \\
\bar{\theta}_3^G
\end{bmatrix}
=\bf{P}_2
\begin{bmatrix}
\bsthetaloc^1\\
\bsthetaloc^2
\end{bmatrix},
\qquad
\bf{P}_2=
\begin{bmatrix}
N_1 & 0 	& 0 	& N_2 	& 0 	& 0 \\
0  	& N_5 	& 0 	& 0 	&N_6 	& 0 \\
0  	&  0  	& N_5 	& 0 	& 0 	& N_6
\end{bmatrix},
\end{equation}
respectively, where $N_1$ and $N_2$ are the linear interpolation functions (for axial displacement) and $N_3$, $N_4$, $N_5$ and $N_6$ are Hermite interpolation functions (for bending).

The position of $G$ in canonical coordinates can be written as:
\begin{equation}\label{eq:preCentroiDisp}
\boldsymbol{OG}    = N_1(\bfx^1+\bfu^1)+N_2(\bfx^2+\bfu^2)+ \bfR_r \bfu_\ell,
\end{equation}
and, using the identities obtained above, the variations of the displacement and rotation of the point $G$ can be written as:
\begin{equation}\label{eq:preDifU}
\delta \bfu = \bfR_r \bfH_1 \bfE^T \delta \bfd_g,
\quad  \text{and} \quad 
\delta \bfw = \bfR_r\bfH_2\bfE^T \delta \bfd_g,
\end{equation}
respectively, where $\bfH_2 = \bfP_2\bfP+\bfG^T$ and $\bfH_1= \bfN+\bfP_1\bfP-\widetilde{\bfu_\ell}\bfG^T$, with $\widetilde{ \bfu_\ell }$ being the skew operator associated with the vector $\bfu_\ell$.

Finally, velocities and accelerations can be obtained as:
\begin{equation} \label{eq:preUdott}
\begin{split}
\dot{\bf{u}} =\bfR_r \bfH_1 \bfE^T \dot{\bfd_g},
\qquad 
\ddot{\bf{u}} = \bfR_r \bfH_1 \bfE^T \ddot{\bfd_g}+\bfR_r \bfC_1 \bfE^T\dot{\bfd_g},
& \\
\dot{\bfw}=\bfR_r\bfH_2\bfE^T\dot{{\bfd_g}},
\quad 
\ddot{\bfw}= \bfR_r\bfH_2\bfE^T\ddot{\bfd_g}+\bfR_r\bfC_2\bfE^T\dot{\bfd_g}.
\end{split}
\end{equation}
where $\bfC_i =  \widetilde{\bfw^e_r}\bfH_i+\dot{\bfH_i}-\bfH_i\bfE_t$ and $\bfw^e_r = \bfG\bfE^T\dot{\bfd_g}$.

\subsection{Internal and inertial forces}

The expressions of the elemental internal and inertial forces in global coordinates can be obtained using the Principle of Virtual Work. Considering Equation~\eqref{eqn:generalPVW} for the internal forces, and substituting the relations presented in Equations~\eqref{eq:preDefR} and \eqref{eq:PreLocalAngVar}, we obtain:
\begin{equation}
 \delta \bfd_g^T	\bff_{g}^{int}  =
	\delta \bfd_g^T	
	\begin{bmatrix}
		\bfr^T & 		\bfE \bfP^T 
	\end{bmatrix} \bff_{\ell}^{int}.
\end{equation}
This identity is valid for any virtual displacement $\delta \bfd_g$, thus we can obtain:
\begin{equation}
	\label{eq:preInternalForce}
	\bff_{g}^{int} =
	\begin{bmatrix}
		\bfr^T & 		\bfE \bfP^T 
	\end{bmatrix} \bff_{\ell}^{int},
\end{equation}
where $\bff_{\ell}^{int}$ is the known vector of internal forces $\bff_{\ell}^{int}=\left[f_{a\ell}~ ~(\bfm_\ell^1)^T~ (\bfm_\ell^2)^T\right]$, with normal force and bending moments, given by a linear constitutive behavior.

For the inertial term, the kinematic energy $K$ of the element in Total Lagrangian coordinates is written as:
\begin{equation}
\label{eq:preKinematicEnergy}
\textit{K}=\frac{1}{2}\int_{l_0} {\rho\dot{\bfu}^T  A \dot{\bf{u}} +
	\rho\dot{\bfw}^T \bfI\dot{\bfw}} ~\text{$dl_0$},
\end{equation}
where $A$ is the area of the cross-section, $\rho$ is the density of the material and $\bfI$ is the geometric inertia tensor. Considering the variation in both members of Equation~\eqref{eq:preKinematicEnergy} and using the derivative chain rule, we obtain: 
\begin{equation}
\label{eq:preDifKinematicEnergy}
\delta\textit{K}=-\int_{l_0} \delta \bfu^T \rho A \ddot{\bfu} + \delta
\bfw^T[\rho\bfI\ddot{\bfw}+\widetilde{\dot{\bfw}}\rho\bfI\ddot{\bfw}]
dl_0.
\end{equation}
The inertial force vector of the element in global coordinates $\bff^{ine}_g$ is then defined consistently by:
\begin{equation}
\label{eq:PreDefFuerzaInercial}
\delta K=(\bff^{ine}_g)^T\delta\bfd_g,
\qquad \text{with} \qquad
\bff^{ine}_g= \int _{l_0} \left \{ \bfH_1^T\bfR_r^T \rho A \ddot{\bfu} +\bfH_2^T \bfR_r [\rho \bfI \ddot{\bfw}+\widetilde{\dot{\bfw}}\rho\bfI\dot{\bfw}] \right \} dl_0.
\end{equation}

\section{Methodology}\label{sec:methodology}

In this section we present the proposed formulation for the computation of the aerodynamic forces and describe a numerical procedure for the resolution of the governing equations.

\subsection{Co-rotational aerodynamic forces}

\begin{figure}[htb]
	\centering
	\def\svgwidth{0.6\textwidth}
	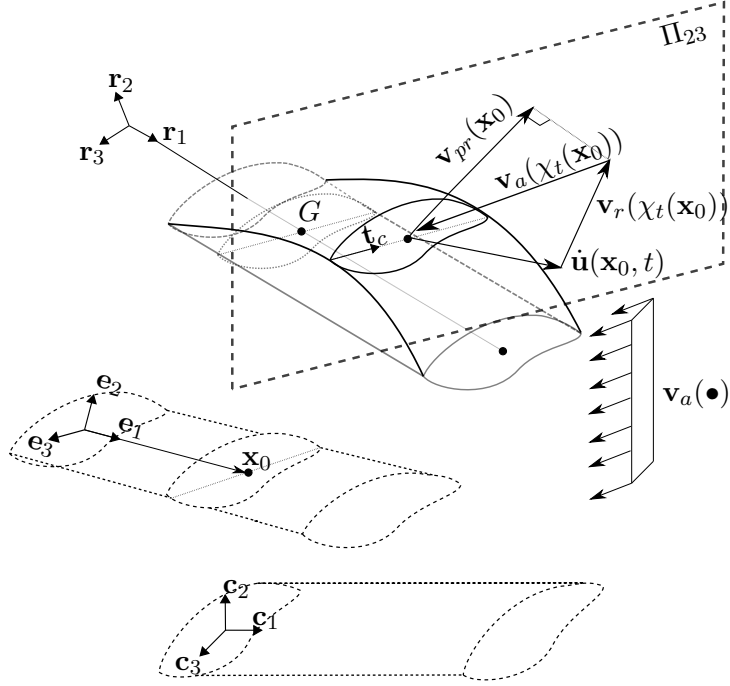
	\caption{Co-rotational framework on fluid loads.}
	\label{fig:velRelProyection}
\end{figure}

Let us consider a frame element, with uniform cross-section, submitted to a fluid flow as shown in Figure~\ref{fig:velRelProyection}. For a section located at $\bfx_0$ with centroid $G$, the deformed position at time $t$ is given by $\bfx=\chi_t(\bfx_0)$. The element is submitted to forces induced by a fluid with absolute velocities given by the field $\bfv_a(\bfx,t):\bbR^3\times \bbR \rightarrow \bbR^3$. The velocity of the centroid is $\dot{\bfu}(\bfx_0,t)$ and the relative velocity in the deformed position is defined by:
\begin{equation}\label{eq:metRelativeVelocity}
\bfv_{r}(\chi_t(\bfx_0),t)=\bfv_{a}(\chi_t(\bfx_0),t)-\dot{\bfu}(\bfx_0,t).
\end{equation}
In this definition a fundamental assumption was considered: the movement of the structure does not affect the absolute velocities of the fluid \citep{Blevins1977}.

\begin{figure}[htb]
		\centering
		\def\svgwidth{0.5\textwidth}
		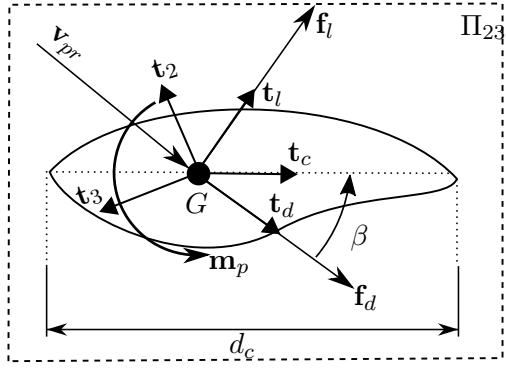
		\caption{Fluid loads on a generic deformed cross-section.}
		\label{fig:dynamicAeroForces}
\end{figure}

The interaction between the fluid flow and the frame element produces normal and shear stresses, that are represented by moments and forces (generalized forces) applied at the deformed position of the centroid. These forces are assumed to be uniquely defined in terms of the instantaneous position and velocity of the deformed section \citep{Foti2018}. In particular, in this formulation, the forces are assumed to depend only on $\bfv_{pr}$ (the projection of the relative velocity onto the plane $\Pi_{23}$ defined by $\bft_2$ and $\bft_3$) as it is shown in Figure~\ref{fig:velRelProyection}. The aerodynamic distributed forces for drag, lift and torsional moment are given by the expressions:
\begin{equation}\label{eq:metAeroForcesGeneral}
\left\{
\begin{array}{ll}
\bff_{d} &= \frac{1}{2}\rho_f d_c c_d(Re,\beta) \, \| \bfv_{pr}  \|^2 \bft_{d},\\
\bff_{l} &=  \frac{1}{2}\rho_f d_c c_l(Re,\beta) \, \| \bfv_{pr} \|^2 \bft_{l},\\
\bfm_{p} & = \frac{1}{2}\rho_f d_c c_m(Re,\beta)\, \| \bfv_{pr} \|^2 \bft_{m},
\end{array}
\right.
\end{equation}
respectively, where $\rho_f$ is the density of the fluid, $d_c$ is the given characteristic dimension of the cross-section and $c_d$, $c_l$ and $c_m$ are the drag, lift and moment coefficients, determined by wind tunnel tests for different Reynolds numbers $Re$ and angles of incidence $\beta$. The angle $\beta$ is defined by $\bfv_{pr}$ and a given unitary vector $\bft_c$, as shown in Figure~\ref{fig:dynamicAeroForces}. It is remarked that drag and lift forces are included in the plane $\Pi_{23}$.

The vector $\bfv_{pr}$ written in the total-deformed system of coordinates is denoted as $(\bfv_{pr})_{\bft}$. In the same manner the notation ($\bullet$)$_{\bft}$ is used for any vector in this system and this sub-index is omitted for vectors in the canonical system. The expression of $(\bfv_{pr})_{\bft}$ is:
\begin{equation}\label{eq:metVprScalar}
	\begin{split}
		(\bfv_{pr})_{\bft}&= (\bfv_{a}-\dot{\bfu})_{\bft}.(\bft_2)_{\bft}+ (\bfv_{a}-\dot{\bfu})_{\bft}.(\bft_3)_{\bft} 
	\end{split}
\end{equation}
where $(\bft_2)_{\bft} = [0, 1, 0]^T$,  $(\bft_3)_{\bft} = [0, 0, 1]^T$ and 
using the rotation matrices of the co-rotational framework as change of basis operators: $\bfR_r = _\bfc\hspace{-0.1cm}(\bfI)_\bfr$, $\overline{\bfR} = _\bfr\hspace{-0.1cm} (\bfI)_\bft$ we can write:
\begin{equation}\label{eq:metVprVector}
	(\bfv_{a}-\dot{\bfu})_{\bft} =\left( \bfR_r  \overline{\bfR}\right) ^T (\bfv_{a}-\dot{\bfu}).
\end{equation}

Substituting Equation~\eqref{eq:metVprScalar} in \eqref{eq:metVprVector} and defining a projection operator $\bfL_2$ we obtain:
\begin{equation}
	(\bfv_{pr})_\bft = \bfL_2 \left( \bfR_r  \overline{\bfR}\right) ^T (\bfv_{a}-\dot{\bfu}).
\end{equation}

Using this we can express the unitary vectors $(\bft_d)_\bft$, $(\bft_l)_\bft$ and $(\bft_m)_\bft$ as:
\begin{eqnarray}
\label{eq:metDragVec}
(\bft_d)_\bft& = & \frac{(\bfv_{pr})_{\bft}}{||(\bfv_{pr})_{\bft}||},\\
\label{eq:metLiftVec}
(\bft_l)_\bft& = & \bfL_3(\bft_d)_\bft ,\\
\label{eq:metMomentVec}
(\bft_m)_\bft & = & (\bft_1)_\bft,
\end{eqnarray}
with $\bfL_3 = \text{exp}([\pi/2, 0, 0]^T)$ and $(\bft_1)_\bft = [1,0,0]^T$. The angle of incidence $\beta$ verifies the identity: 
\begin{equation}\label{eq:metCosBeta}
 (\bft_d)_\bft \cdot (\bft_c)_\bft = \| (\bft_d)_\bft \| \| (\bft_c)_\bft \| \cos(\beta),
\end{equation}
and considering that $\bft_d$ and $\bft_c$ are unitary we obtain the expression:
\begin{equation}\label{eq:angOfAttack}
\beta = \text{sign}
\left[
  \left(
    (\bft_d)_\bft \wedge (\bft_c)_\bft  
  \right)
  \cdot (\bft_1)_\bft
\right] . \arccos((\bft_d)_\bft.(\bft_c)_\bft), 
\end{equation}
where a convention was considered as shown in Figure~\ref{fig:dynamicAeroForces}.
Once the value of $\beta$ is computed at a certain $Re$, the values of $c_d$, $c_l$ and $c_m$ can be determinate using wind tunnel results for the specific cross-section.

Substituting the identities obtained above in Equation~\eqref{eq:metAeroForcesGeneral} we obtain:
\begin{equation}\label{eqn:metAeroForcesCorot}
\begin{array}{lll}
(\bff_{d})_\bft &=&  \frac{1}{2}\rho_fd_c c_d ||\bfL_2 \left( \bfR_r  \overline{\bfR}\right)^T  (\bfv_{a}-\dot{\bfu})||\bfL_2 \left( \bfR_r  \overline{\bfR}\right)^T (\bfv_{a}-\dot{\bfu}), \\
(\bff_{l})_\bft &=& \frac{1}{2}\rho_fd_c c_l ||\bfL_2\left( \bfR_r  \overline{\bfR}\right)^T (\bfv_{a}-\dot{\bfu})||\bfL_3\bfL_2 \left( \bfR_r  \overline{\bfR}\right)^T (\bfv_{a}-\dot{\bfu}),\\
(\bfm_{p})_\bft &=& \frac{1}{2}\rho_fd_c c_m ||\bfL_2\left( \bfR_r  \overline{\bfR}\right)^T (\bfv_{a}-\dot{\bfu})||^2 . (\bft_1)_\bft.
\end{array}
\end{equation}

The virtual work corresponding to the aerodynamic forces of the element is:
\begin{equation}\label{eq:aeroElementVirtualWork}
\delta W_{f} =\int_{l_0} \left\{\delta \bfu^T   \bfR_r\overline{\bfR} (\bff_{d+l})_\bft + \delta \bfw^T \bfR_r\overline{\bfR}(\bfm_p)_\bft \right\}dl_0. 
\end{equation}
where $\bff_{d+l}$ is the sum of $\bff_d$ and $\bff_l$. Considering the vector of nodal aerodynamic generalized forces in global coordinates $\bff^{flu}_{g}$, the virtual work can also be written as: 
\begin{equation}\label{eq:aeroNodalVirtualWork}
	\delta W_{f} =(\delta \bfd_g)^T ~.~\bff^{flu}_{g},
\end{equation}
substituting Equation~\eqref{eq:preDifU} in \eqref{eq:aeroElementVirtualWork} and using Equation~\eqref{eq:aeroNodalVirtualWork} we obtain:
\begin{equation}\label{eq:virtualWorkEquivalentForceExpression}
 (\delta \bfd_g)^T \bff^{flu}_{g}  =\int_{l_0}\left\{\delta \bfd_g^T \bfE \bfH_1 ^T \bfR_r^T\bfR_r \overline{\bfR} (\bff_{d+l})_\bft+ \delta \bfd_g^T \bfE \bfH_2^T \bfR_r^T\bfR_r\overline{\bfR}(\bfm_{p})_\bft  \right\}  dl_0.
\end{equation}
Operating we obtain:
\begin{equation}\label{eq:fagGlobalGeneric}
\bff^{flu}_{g} = 
\bfE 
\left[ \int_{l_0} \left\{ \bfH_1 ^T\overline{\bfR}  (\bff_{l+d})_\bft+ \bfH_2^T\overline{\bfR} (\bfm_{p})_\bft  \right\}  dl_0 \right],
\end{equation}
and substituting Equation~\eqref{eqn:metAeroForcesCorot} the complete expression of the aerodynamic forces vector is obtained:
\begin{eqnarray}
\bff^{flu}_{g}  &=&\frac{1}{2}\rho_f d_c \bfE 
\left( 
\int_{l_0} \left\{\bfH_1 ^T  \overline{\bfR}||\bfL_2 \left( \bfR_r  \overline{\bfR}\right)^T  (\bfv_{a}-\dot{\bfu})|| 
\begin{bmatrix} 
(c_d\bfI + c_l\bfL_3 ) 
\end{bmatrix}
\bfL_2 \left( \bfR_r  \overline{\bfR}\right)^T (\bfv_{a}-\dot{\bfu})\right\} dl_0 \right. \dots
\nonumber \\
& \dots & +
\left. \int_{l_0} 
\left\{
  \bfH_2^T\overline{\bfR} c_m ||\bfL_2 
  \left( \bfR_r  \overline{\bfR}
  \right)^T (\bfv_{a}-\dot{\bfu})||^2 (\bft_1)_\bft
\right\}  dl_0
\right)
\label{eq:metFinalAeroForce}
\end{eqnarray}

\subsection{Balance equations and numerical resolution procedure}

The governing equations are obtained by considering the virtual work for all the elements of the structure  for the forces in Equations~\eqref{eq:preInternalForce}, \eqref{eq:PreDefFuerzaInercial} and \eqref{eq:metFinalAeroForce}. Additionally a vector with external forces not induced by the fluid interaction $\bff^{ext}_{g,s}$. The nonlinear system of equations is written as:
\begin{equation}
	\bff^{res}=  \bfzero, \quad \text{with } \quad \bff^{res}= \bff^{ext}_{g,s}(t) + \bff^{flu}_{g,s}(\bfd_{g,s}, \dot{\bfd}_{g,s}) - \bff^{int}_{g,s}(\bfd_{g,s}) - \bff^{ine}_{g,s}(\bfd_{g,s}, \dot{\bfd}_{g,s}, \ddot{\bfd}_{g,s}).
\end{equation}
where the arguments of the residual forces $\bff^{res}$ were omitted.

The numerical resolution procedure proposed consists in solving the system of nonlinear governing equations using iterative methods. Two reference numerical methods were implemented: the Newmark method with $\alpha_N=1/4$, $\delta_N=1/2$ and the $\alpha$-HHT method with $\alpha_H=-0.05$ \citep{Bathe2019}. 

The approach proposed for the computation of the tangent matrices of the methods consists in neglecting the tangent matrix for the aerodynamic forces and consider the reference literature for the internal $\bff^{int}_g$ and inertial $\bff^{ine}_g$ forces \citep{Le2014,Battini2002}. 

The computation of the aerodynamic and inertial force vector of the element is done using the integration Gauss method.  

\section{Numerical results}\label{sec:numericalResults}

In this section the numerical results obtained for five problems are presented. For all the problems the fluid considered is air with density $\rho_f = 1.225$ kg/m$^3$, kinematic viscosity $\nu_f = 1.5\times10^{-5}$~m$^2$/s, at 20 $^\circ$C and atmospheric pressure. Regarding the elastic properties, Poisson's ratio $\nu=0.3$ is considered for the first four examples. For all the problems, homogeneous initial conditions are considered.

All the numerical results presented can be reproduced by running the scripts publicly available. The results shown were produced using a computer with a Linux OS, a 64-bit architecture, an Intel i7-6700HQ CPU and 8 Gb of RAM, running the implementation of the formulation in ONSAS on GNU-Octave \citep{Eaton2015a}. The visualization is done using Paraview \citep{Geveci2005} and GNU-Octave. 

The stopping criteria considered in all the examples are given by:
\begin{equation}
\frac{ \| \Delta \bfd_{g,s}^k\| }{ \| \bfd_{g,s}^k\| }  \leq tol_u  \quad \text{and} \quad \| \Delta \bff^{res,k} \| \leq tol_r ,
\end{equation}
where $k$ is the number of iteration and $tol_u$, $tol_r$ are scalars to be defined. These criteria are implemented within ONSAS.

\subsection{Example 1: cylindrical cantilever beam submitted to small-displacements}\label{sec:cylindricalCantBeam}

In this example a simple cantilever problem with semi-analytic solution is considered. The main goal is to validate the formulation, and verify its numerical implementation, for a small-displacements case. 

\newcommand{\SDCantBeamL}{5}
\newcommand{\SDCantBeamd}{0.1}
\newcommand{\SDCantBeamE}{5}
\newcommand{\SDCantBeamnu}{0.3}
\newcommand{\SDCantBeamrho}{700}
\newcommand{\SDCantBeamcd}{1.2}
\newcommand{\SDCantBeamva}{15}
\newcommand{\SDCantBeamtolF}{1}
\newcommand{\SDCantBeamtolU}{1}
\newcommand{\SDCantBeamdeltaT}{0.05}
\newcommand{\SDCantBeamfinalTime}{20}
\newcommand{\SDCantBeamnumElem}{10}

\subsubsection{Problem definition}\label{subsecCylindircalCantBeamProblemDef}

\begin{figure}[ht]
	\begin{subfigure}{0.42\textwidth}
		\def\svgwidth{0.86\textwidth}
		\centering
		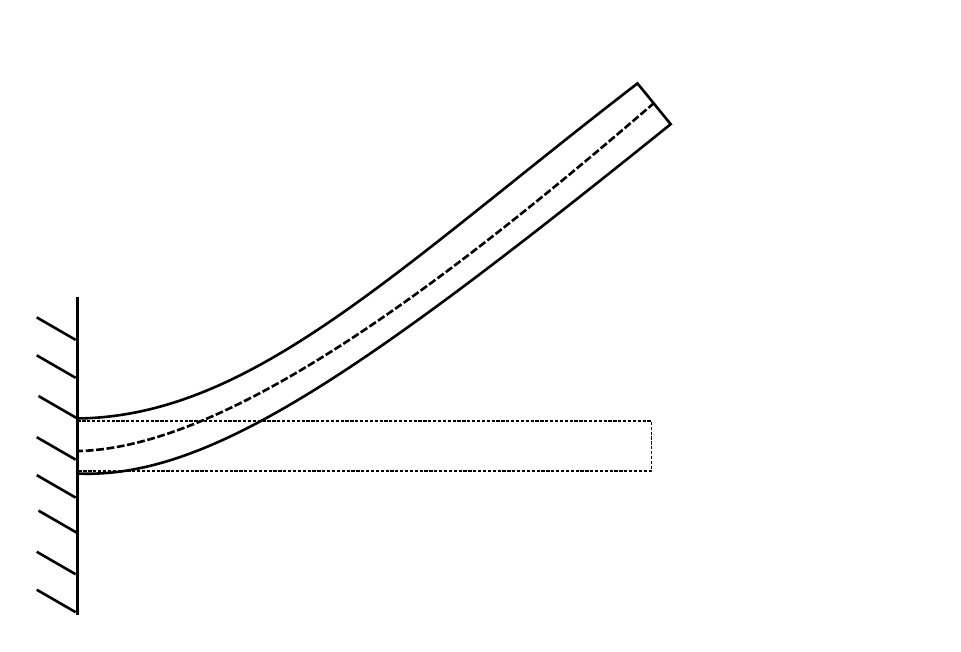
		\caption{Geometry of cantilever beam with boundary conditions and flow.}
		\label{fig:cylindricalCantBeamIlus}
	\end{subfigure}
	\begin{subfigure}{.55\textwidth}
		\def\svgwidth{0.9\textwidth}
		\centering
		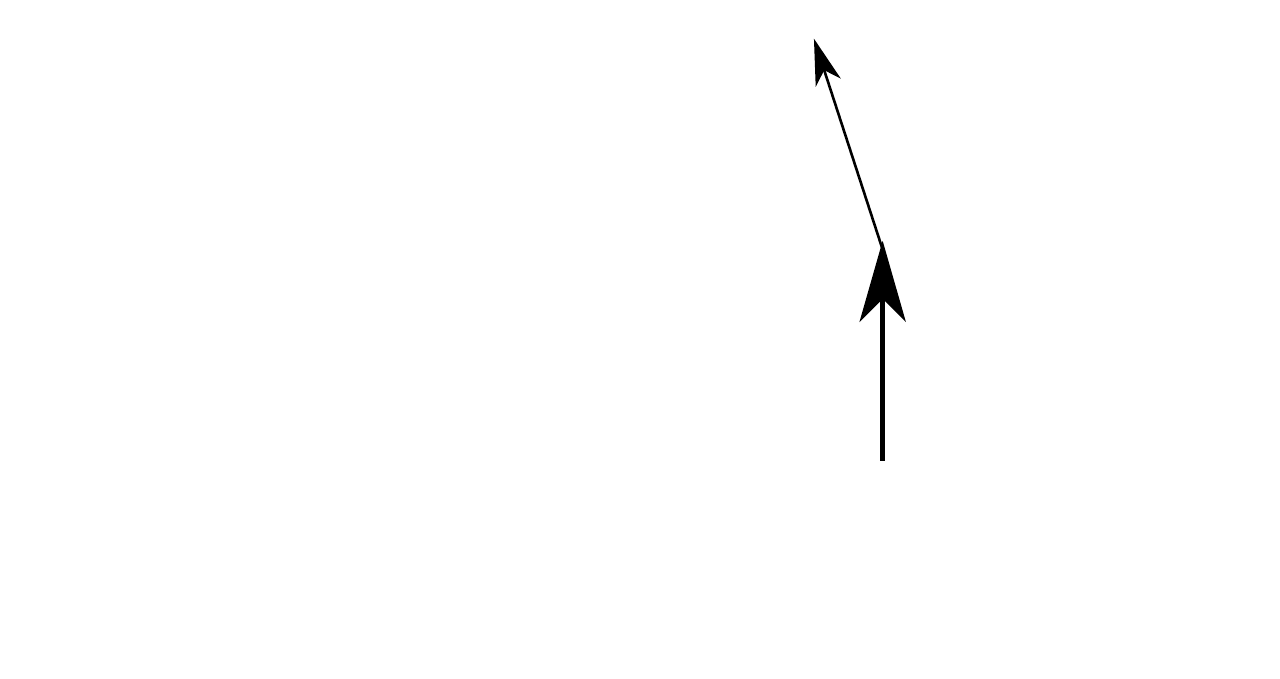
		\caption{Absolute and projected transversal velocities for the steady case.}
		\label{fig:cylindricalCantBeamProyection}
	\end{subfigure}
	\caption{Example 1: Diagram of cantilever beam and wind flow.}
	\label{fig:cylindricalCantBeamData}
\end{figure}

The problem consists in a cantilever beam submitted to a uniform wind field $\bfv_a(\bfx,t) = v_a(t) \bfc_2$ as it is shown in Figure~\ref{fig:cylindricalCantBeamIlus}. The beam is clamped on the boundary $x=0$ m, free on the node A, and the span length is $L = \SDCantBeamL$ m. The cross-section is circular with diameter $d = \SDCantBeamd$ m. The chord used to compute the aerodynamic forces from Equation~\eqref{eq:metFinalAeroForce} is $d_c=d$. A material with Young modulus $E=\SDCantBeamE$ GPa and density $\rho = \SDCantBeamrho$ kg/m$^3$ is considered. The behavior of the beam in this case is assumed linear elastic, with small displacements and small rotations.

The aerodynamic coefficients are taken from \citep{Roshko1961}, assuming a sub-critical flow with a Reynolds number $Re = \frac{v_ad}{\nu_f} =10^5$, therefore we obtain $c_d=\SDCantBeamcd$ and $c_l=c_m=0$. The velocity corresponding to this $Re$ is $v_a=\SDCantBeamva$ m/s.

\subsubsection{Numerical results: steady case}\label{subsecCylindircalCantBeamStaticSmallDisp}

The goal of this analysis case is to obtain the deformed configuration of the beam at the steady state, both numerically and analytically. The flow velocity assumed is $v_a(t) = \SDCantBeamva$ m/s, and since this is a steady case, the velocity of any point of the beam is $\dot{\bfu}=0$. Substituting this in Equation~\eqref{eq:metRelativeVelocity} we obtain $\bfv_r(x,t) = \bfv_a(x,t)$. In Figure~\ref{fig:cylindricalCantBeamProyection} the absolute velocity of the fluid and the relative projected transversal component of the velocity are shown, and for this analysis case, we obtain the identity:
\begin{equation}\label{eqn:projej1}
	\left\| \bfv_{pr}(x) \right\| = \| \bfv_{a}(x) \| |\cos\left( \theta_z(x)\right)  | .
\end{equation} 
The distributed drag force in the $y$ direction is obtained with Equation~\eqref{eq:metAeroForcesGeneral} and using the Euler-Bernoulli static beam equation we obtain:
\begin{equation}\label{eq:resCylindricalCantBeamDiffEq3order}
EI_{zz} \frac{\partial ^3 \theta_z}{\partial x ^3} = -q_0 c_d\cos^3(\theta_z), 
\end{equation}
where $q_0 = \frac{1}{2}\rho_f d_c v_a^2$ and $I_{zz}$ is the second moment of inertia. %
This nonlinear equation is solved using the Julia library DifferentialEquations.jl \citep{Rackauckas2017}, providing a \textit{semi-analytic} solution that is used to verify the numerical solutions.

To obtain the numerical solutions for the steady case, a nonlinear static problem is solved, using a Newton-Raphson method. For the spatial discretization, $\SDCantBeamnumElem$ elements are considered and for the numerical integration of the aerodynamic force vector, 4 Gauss integration points are used. The stopping criteria for the numerical iterative method are $tol_r=\SDCantBeamtolF0^{-10}$ and $tol_u=\SDCantBeamtolU0^{-10}$.

\begin{figure}[htb]
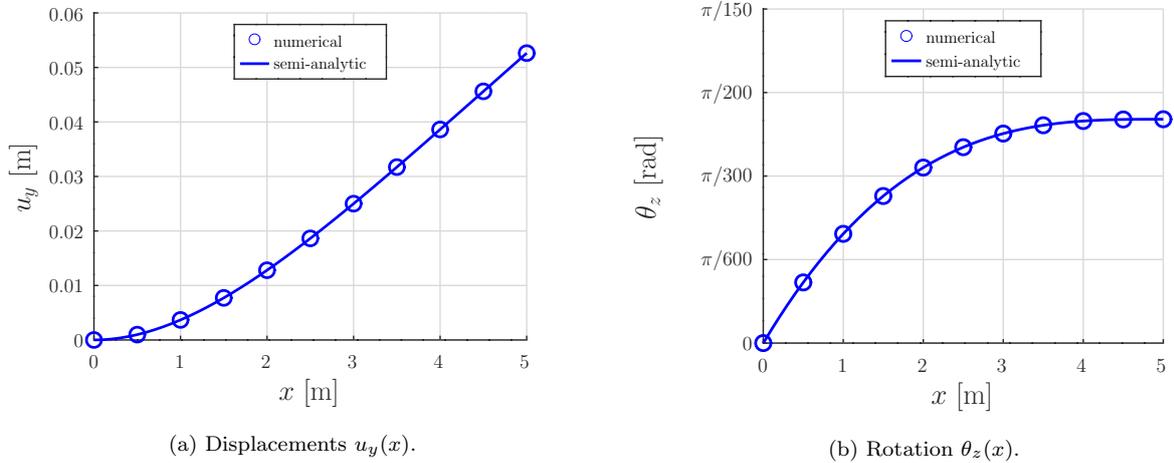

	\begin{subfigure}{.5\textwidth}
	\centering
	\resizebox{.93\textwidth}{!}{\input{CilCantSDLinDisp.tex}}
	\caption{Displacements $u_y(x)$.}
	\label{fig:cylindricalCantBeamLinDisp}
\end{subfigure}
	\begin{subfigure}{0.5\textwidth}
		\centering
		\resizebox{.95\textwidth}{!}{\input{CilCantSDAngDisp.tex}}
		\caption{Rotation $\theta_z(x)$.}
		\label{fig:cylindricalCantBeamAngDisp}
	\end{subfigure}
	\caption{Example 1: Numerical results obtained for the steady case and comparison with semi-analytic solution.}
	\label{fig:cylindricalCantBeamValidDisp}
\end{figure}

The results obtained for $\theta_z$ and $u_y$  are shown in Figure~\ref{fig:cylindricalCantBeamValidDisp}, where the circles represent the numerical solution at each node.
It is observed that the numerical results match the semi-analytic solution, allowing us to conclude that the formulation and the implementation are validated for this case.

\subsubsection{Numerical results: non-steady case}

In this section the numerical solutions of the problem for non-steady (or time-varying) flows are presented.
In this case the wind velocity $v_a$ is given by:
\begin{equation}
	v_a(x,t)=
	\left\{
	\begin{array}{lr}
\displaystyle	\frac{t}{t_c} \, \SDCantBeamva \text{ m/s} & t \in [0, t_c], \\
	\SDCantBeamva \text{ m/s} 			   & t \in (t_c,+\infty),  \\
\end{array}
\right.
\end{equation}
where two values are considered for $t_c$: $t_c=1$ s (case 1), and $t_c=4$ s (case 2). The trapezoidal Newmark numerical method is used, with time step $\Delta t = \SDCantBeamdeltaT$ s and the same stopping criteria considered in the steady case. For the inertial terms the consistent mass matrix of the small-displacement Euler-Bernoulli beam element is considered \citep{Chuhan2014}.

\begin{figure}[htb]
	\centering
	\resizebox{.6\textwidth}{!}{\input{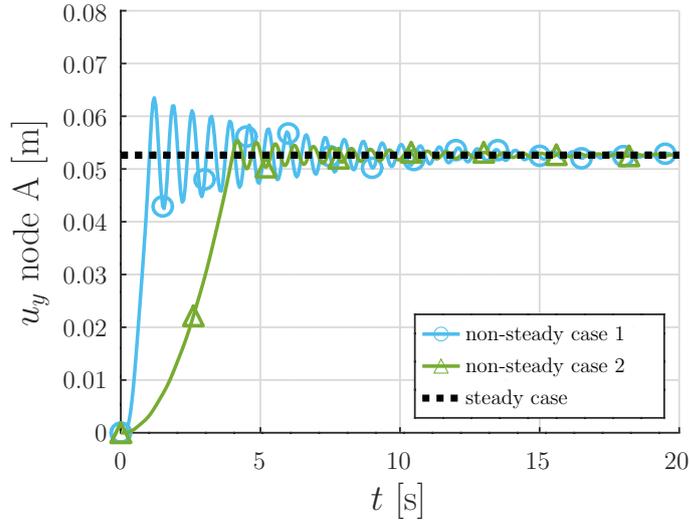}}
	\caption{Example 1: Evolution of $u_y$ displacement of node A.}
	\label{fig:cylindricalCantvUyA}
\end{figure}

The solutions obtained for the displacement $u_y$ of point A are shown in Figure~\ref{fig:cylindricalCantvUyA}. It can be observed that both non-steady solutions converge to the steady one, providing the expected behavior and validating the proposed formulation and implementation, for these cases.
 
\clearpage
\subsection{Example 2: cylindrical cantilever beam submitted to large-displacements}
\newcommand{\LDCantBeamL}{10}
\newcommand{\LDCantBeamd}{0.2}
\newcommand{\LDCantBeamE}{1}
\newcommand{\LDCantBeamrho}{300}
\newcommand{\LDCantBeamcd}{1.2}
\newcommand{\LDCantBeamva}{0}
\newcommand{\LDCantBeamtolF}{1}
\newcommand{\LDCantBeamtolU}{0}
\newcommand{\LDCantBeamdeltaT}{0.2}
\newcommand{\LDCantBeamfinalTime}{150}
\newcommand{\LDCantBeamtimeError}{65}

The main goal of this example is to verify the formulation for a problem with large-displacement motions and oscillating fluid velocity.

\subsubsection{Problem definition}
\begin{figure}[htb]
	\begin{subfigure}{.47\textwidth}
		\centering
		\def\svgwidth{0.95\textwidth}
		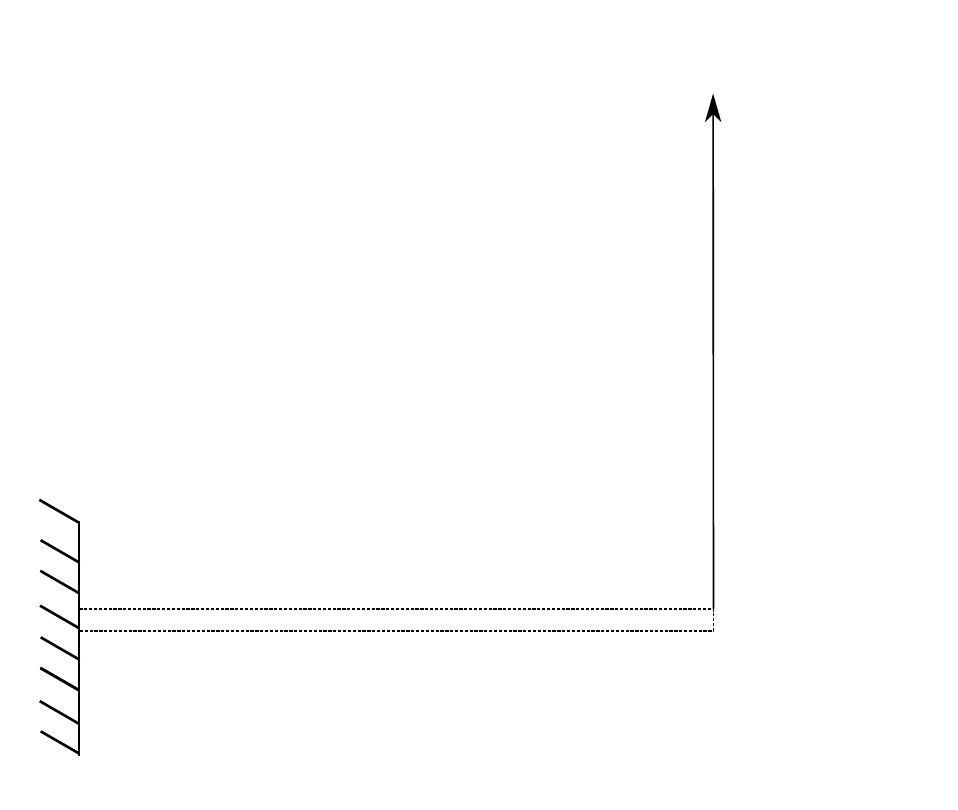
		\caption{Diagram of geometry and boundary conditions.}
		\label{fig:LDcylindricalCantBeamIlus}
	\end{subfigure}
	\begin{subfigure}{.52\textwidth}
		\centering
		\resizebox{.95\textwidth}{!}{\input{CilCantLDWindProfile.tex}}
		\caption{Absolute wind velocity profile.}
		\label{fig:cylindricalCantwindLDWindVel}
	\end{subfigure}
	\caption{Example 2: Diagram of cantilever beam and wind velocity profile.}
\end{figure}

The problem consists in a cantilever beam with solid circular cross-section of diameter $d=\LDCantBeamd$~m and a span length $L=\LDCantBeamL$ m, submitted to a flow as shown in Figure~\ref{fig:LDcylindricalCantBeamIlus}. The wind velocity considered is $\bfv_a(\bfx,t) = v_a(t) \bfc_2$ where the function $v_a$ is shown in Figure~\ref{fig:cylindricalCantwindLDWindVel}, and can be written as:
\begin{equation}\label{eq:resLDCylindricalCantBeamWindVel}
	v_a(t)=
	\left\{
	\begin{array}{lr}
		\left( 5 \sin(\omega_a t)  + \sin(\omega_b t) \right)  ~\text{m/s} & t \in[0, 50 \text{ s}] \\
		0 ~\text{m/s}			   & t \in (50 \text{ s}, +\infty) \\
	\end{array}
\right.
\end{equation}
with $\omega_a = 2\pi/50$ rad/s and $ \omega_b=2\pi/0.77$ rad/s.
The drag coefficient is assumed constant $c_d=1.2$, according to  Wieselsberger’s curve \citep{Roshko1961}, and $c_l = c_m=0$.

The constitutive behavior is considered linear (with the engineering strain) with Young modulus $E = \LDCantBeamE$ MPa. Geometric nonlinearity is considered within the co-rotational frame element formulation. For the inertial terms both lumped-mass and consistent formulations are implemented and the density considered is $\rho = \LDCantBeamrho$ kg /m$^3$.

\subsubsection{Numerical results}

The $\alpha$-HHT numerical time integration method with time step $\Delta_t = \LDCantBeamdeltaT$ s is used for all the cases. The stopping criteria in this case is set only by the norm of the residual force $tol_r=\LDCantBeamtolF0^{-8}$.

\begin{figure}[htb]
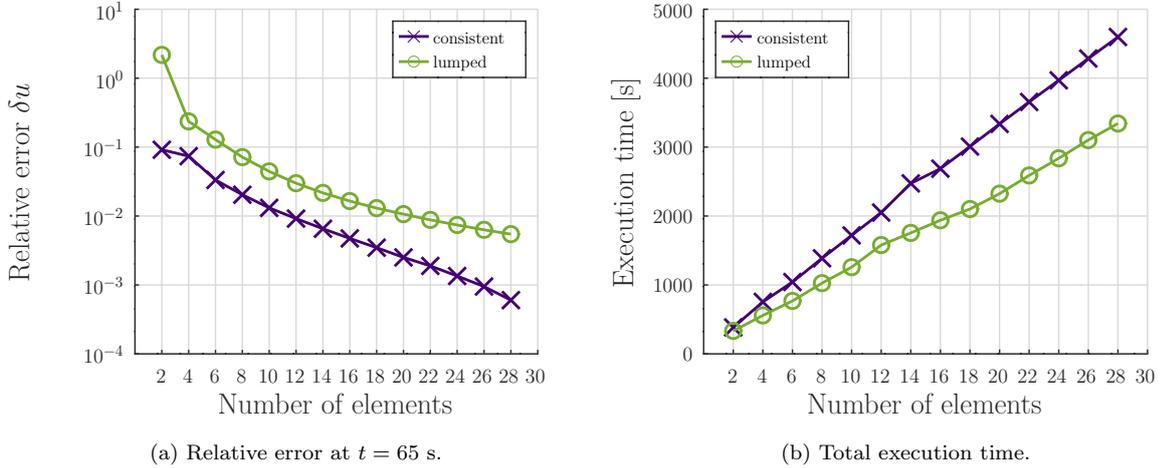

	\begin{subfigure}{.48\textwidth}
		\centering
		\resizebox{\textwidth}{!}{\input{CilCantLDMeshAnalysis.tex}}
		\caption{ Relative error at $t=\LDCantBeamtimeError$ s.}
		\label{fig:cylindricalCantvLDMeshAnalysis}
	\end{subfigure}
	\begin{subfigure}{.48\textwidth}
		\centering
		\resizebox{\textwidth}{!}{\input{CilCantLDExecTime.tex}}	
		\caption{Total execution time.}
		\label{fig:cylindricalCantvLDExecutionTime}
	\end{subfigure}
	\caption{Example 2: Mesh analysis and execution time results.}
	\label{fig:cylindricalCantvLDAnalysis}
\end{figure}

A mesh analysis is performed, where numerical solutions for time $t=\LDCantBeamtimeError $ s are compared. The reference solution $\bfu_{ref}(x_0,t)$ is obtained using 30 elements with the consistent inertial formulation. The relative difference between displacement fields is computed considering the following metric: 
\begin{equation} \label{eq:resReleativeErrorLDcantBeam}
\delta_u = \frac{ \int_{\ell_0} \left| \bfu(x_0,t) - \bfu_{ref}(x_0,t) \right| d x_0 }{ \int_{\ell_0} \left| \bfu_{ref}(x_0) \right| d x_0 }.
\end{equation}
The results obtained are shown in Figure~\ref{fig:cylindricalCantvLDMeshAnalysis}, while total execution time for different cases are shown in Figure~\ref{fig:cylindricalCantvLDExecutionTime}. %

\begin{figure}[htb]
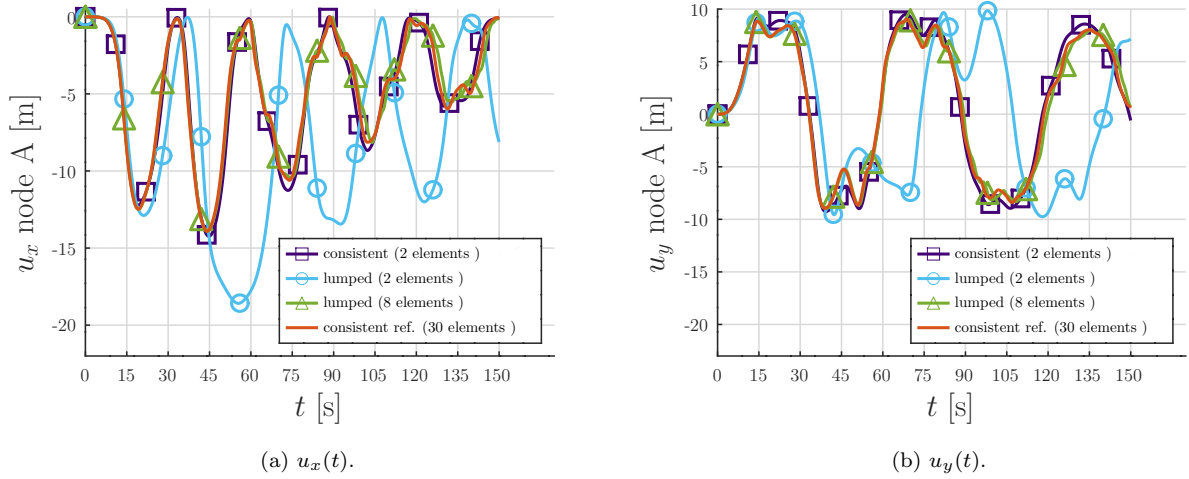

	\begin{subfigure}{0.5\textwidth}
		\centering
		\resizebox{.98\textwidth}{!}{\input{CilCantLDUxA.tex}}
		\caption{$u_x(t)$.}
		\label{fig:cylindricalCantvLDDispsUX}
	\end{subfigure}
	\begin{subfigure}{.5\textwidth}
		\centering
		\resizebox{.98\textwidth}{!}{\input{CilCantLDUyA.tex}}		
		\caption{$u_y(t)$.}
		\label{fig:cylindricalCantvLDDispsUY}
	\end{subfigure}
	\caption{Example 2: Displacements $u_x(t)$ and $u_y(t)$ of node A.}
	\label{fig:cylindricalCantwindLDDisps}
\end{figure}

The  results obtained for the consistent formulation (using 2 and 30 elements) and for the lumped formulation (using 2 and 8 elements) are shown in Figure~\ref{fig:cylindricalCantwindLDDisps}. %

\begin{figure}[htb]
	\centering
	\def\svgwidth{0.6\textwidth}
\begingroup%
  \makeatletter%
  \providecommand\color[2][]{%
    \errmessage{(Inkscape) Color is used for the text in Inkscape, but the package 'color.sty' is not loaded}%
    \renewcommand\color[2][]{}%
  }%
  \providecommand\transparent[1]{%
    \errmessage{(Inkscape) Transparency is used (non-zero) for the text in Inkscape, but the package 'transparent.sty' is not loaded}%
    \renewcommand\transparent[1]{}%
  }%
  \providecommand\rotatebox[2]{#2}%
  \newcommand*\fsize{\dimexpr\f@size pt\relax}%
  \newcommand*\lineheight[1]{\fontsize{\fsize}{#1\fsize}\selectfont}%
  \ifx\svgwidth\undefined%
    \setlength{\unitlength}{760.48113722bp}%
    \ifx\svgscale\undefined%
      \relax%
    \else%
      \setlength{\unitlength}{\unitlength * \real{\svgscale}}%
    \fi%
  \else%
    \setlength{\unitlength}{\svgwidth}%
  \fi%
  \global\let\svgwidth\undefined%
  \global\let\svgscale\undefined%
  \makeatother%
  \begin{picture}(1,0.6666832)%
    \lineheight{1}%
    \setlength\tabcolsep{0pt}%
    \put(0,0){\includegraphics[width=\unitlength,page=1]{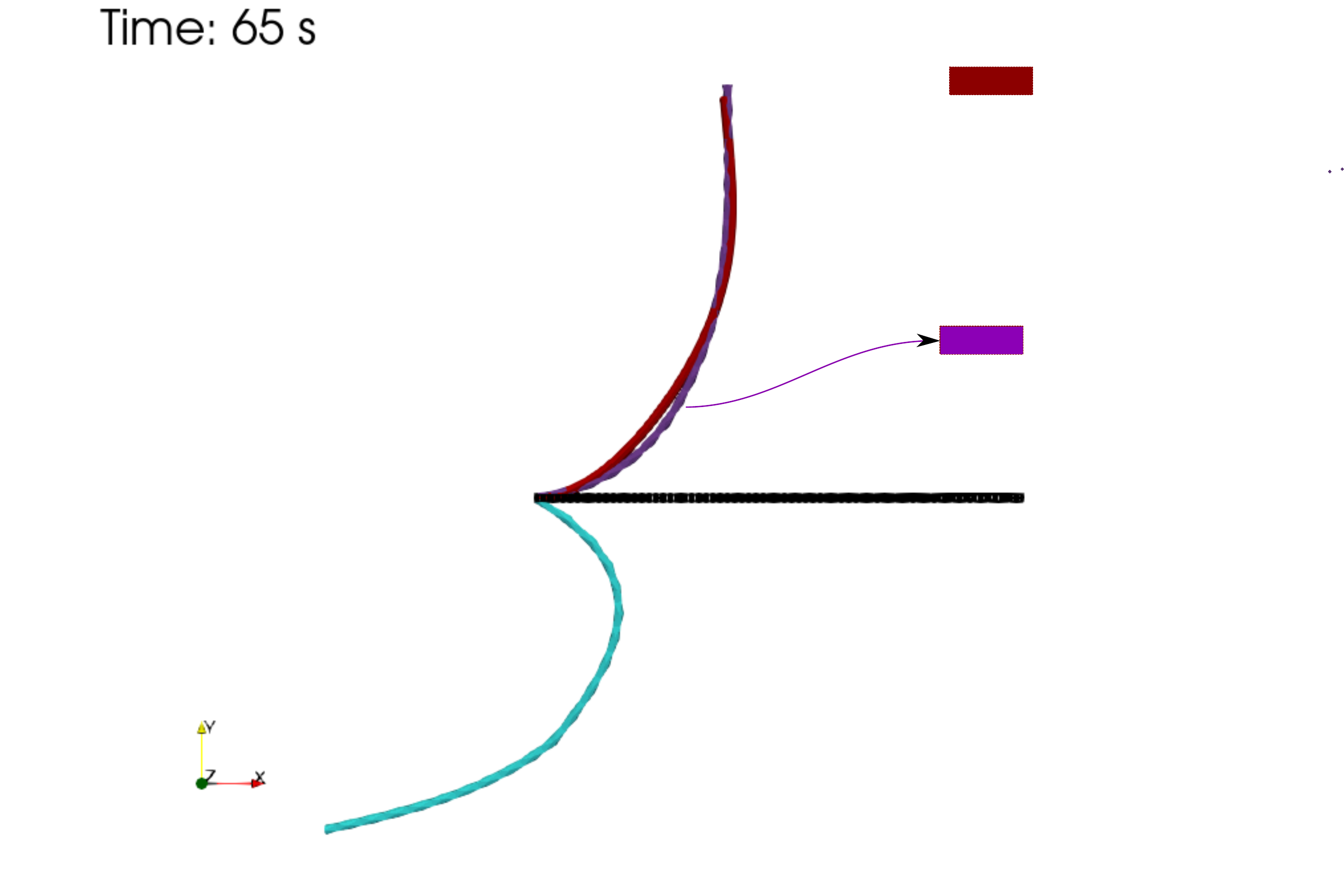}}%
    \put(0.77758766,0.60025932){\makebox(0,0)[lt]{\lineheight{1.25}\smash{\begin{tabular}[t]{l}consistent ref.\end{tabular}}}}%
    \put(0.76391496,0.40532228){\makebox(0,0)[lt]{\lineheight{1.25}\smash{\begin{tabular}[t]{l}consistent 2 elem.\end{tabular}}}}%
    \put(0.69290602,0.19690791){\makebox(0,0)[lt]{\lineheight{1.25}\smash{\begin{tabular}[t]{l}($t=0$ s) \end{tabular}}}}%
    \put(0,0){\includegraphics[width=\unitlength,page=2]{defLD.pdf}}%
    \put(-0.00178584,0.21561974){\makebox(0,0)[lt]{\lineheight{1.25}\smash{\begin{tabular}[t]{l}lumped 2 elem.\end{tabular}}}}%
    \put(0,0){\includegraphics[width=\unitlength,page=3]{defLD.pdf}}%
  \end{picture}%
\endgroup%

	\caption{Example 2: Consistent and lumped with 2 elements, and reference deformed configurations at $t=\LDCantBeamtimeError$ s.}
	\label{fig:cylindricalCantvLDDef}
\end{figure}

The deformed configurations at time $t=\LDCantBeamtimeError$ s, for the lumped and consistent formulations with 2 elements are shown in Figure~\ref{fig:cylindricalCantvLDDef}. %
It can be seen that the lumped formulation diverges from the reference solution at an early stage of the simulation.

The results obtained let us conclude that the proposed formulation integrating drag and consistent inertial forces provides the expected behavior of the problem. Moreover, it is confirmed that the consistent formulation is more efficient than the lumped formulation, particularly for problems with large displacements and rapid fluid velocity variations.

\clearpage
\subsection{Example 3: Cantilever beam with square cross-section}\label{sec:squareCantBeam}

\newcommand{\SquareCantBeamL}{2}
\newcommand{\SquareCantBeama}{0.2}
\newcommand{\SquareCantBeamE}{35}
\newcommand{\SquareCantBeamrho}{800}
\newcommand{\SquareCantBeamvm}{2.8}
\newcommand{\SquareCantBeamtolF}{1}
\newcommand{\SquareCantBeamtolU}{1}
\newcommand{\SquareCantBeamdeltaT}{0.5}
\newcommand{\SquareCantBeamfinalTime}{30}
\newcommand{\SquareCantBeamnumElem}{10}
\newcommand{\SquareCantBeamtimeError}{8}

In this example a cantilever beam with a square cross-section is considered, where the drag and lift aerodynamic coefficients are given by realistic functions for any angle of incidence $\beta$. The goal of the example is to verify the formulation and the numerical iterative approach considered for the resolution of the balance equations. The variation of the results for different numbers of Gauss integration points is also studied.

\subsubsection{Problem definition}\label{sec:squareCantBeamPD}

\begin{figure}[htb]
	\begin{subfigure}{0.46\textwidth}
	\def\svgwidth{0.98\textwidth}
	\centering
	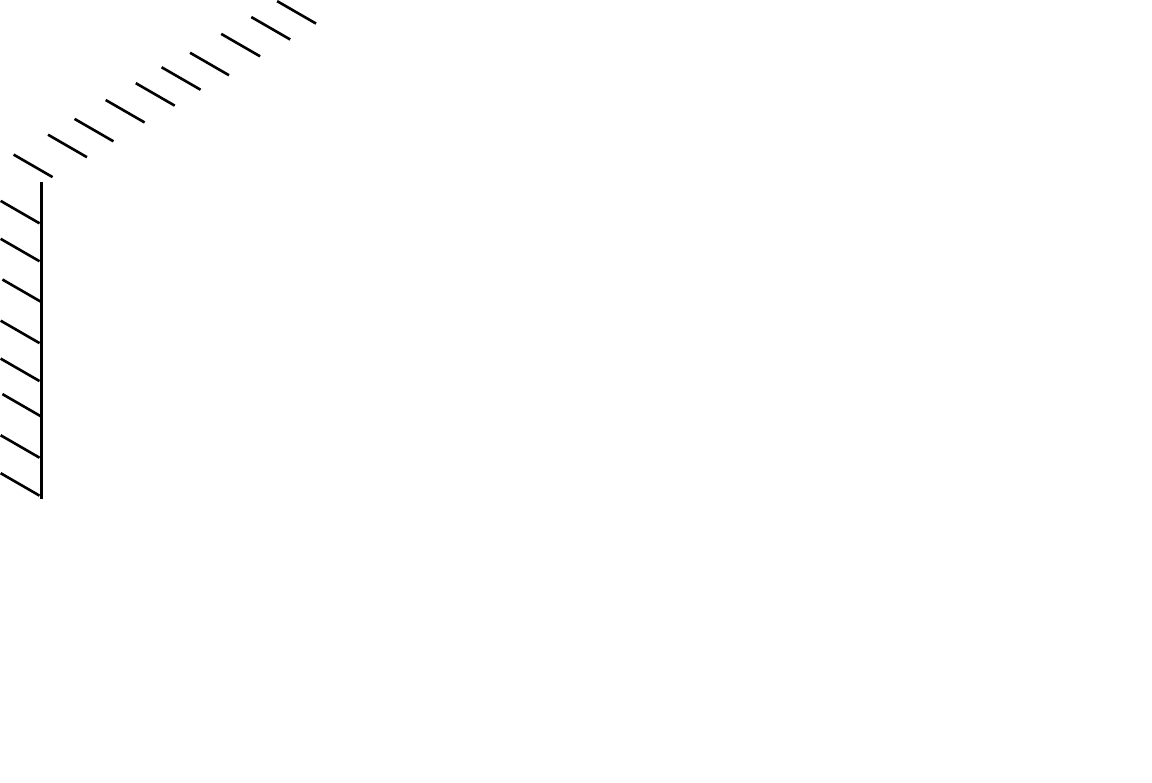
	\caption{Square cantilever beam general layout.}
	\label{fig:squareCantBeamIlus}
	\end{subfigure}
	\begin{subfigure}{.51\textwidth}
		\def\svgwidth{0.6\textwidth}
		\centering
\begingroup%
  \makeatletter%
  \providecommand\color[2][]{%
    \errmessage{(Inkscape) Color is used for the text in Inkscape, but the package 'color.sty' is not loaded}%
    \renewcommand\color[2][]{}%
  }%
  \providecommand\transparent[1]{%
    \errmessage{(Inkscape) Transparency is used (non-zero) for the text in Inkscape, but the package 'transparent.sty' is not loaded}%
    \renewcommand\transparent[1]{}%
  }%
  \providecommand\rotatebox[2]{#2}%
  \newcommand*\fsize{\dimexpr\f@size pt\relax}%
  \newcommand*\lineheight[1]{\fontsize{\fsize}{#1\fsize}\selectfont}%
  \ifx\svgwidth\undefined%
    \setlength{\unitlength}{70.49597956bp}%
    \ifx\svgscale\undefined%
      \relax%
    \else%
      \setlength{\unitlength}{\unitlength * \real{\svgscale}}%
    \fi%
  \else%
    \setlength{\unitlength}{\svgwidth}%
  \fi%
  \global\let\svgwidth\undefined%
  \global\let\svgscale\undefined%
  \makeatother%
  \begin{picture}(1,1.05995997)%
    \lineheight{1}%
    \setlength\tabcolsep{0pt}%
    \put(0.32626917,0.33540529){\color[rgb]{0,0,0}\rotatebox{-0.46513754}{\makebox(0,0)[lt]{\lineheight{1.25}\smash{\begin{tabular}[t]{l}$\bft_1$\end{tabular}}}}}%
    \put(0.451023,0.79441319){\color[rgb]{0,0,0}\makebox(0,0)[lt]{\lineheight{1.25}\smash{\begin{tabular}[t]{l}$\bft_3$\end{tabular}}}}%
    \put(0.54714796,0.52514353){\color[rgb]{0,0,0}\makebox(0,0)[lt]{\lineheight{1.25}\smash{\begin{tabular}[t]{l}$\bft_2=\bft_c$\end{tabular}}}}%
    \put(0.61449989,0.11881273){\color[rgb]{0,0,0}\makebox(0,0)[lt]{\lineheight{1.25}\smash{\begin{tabular}[t]{l}$\bft_{d}$\end{tabular}}}}%
    \put(0,0){\includegraphics[width=\unitlength,page=1]{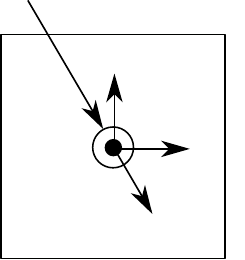}}%
    \put(0.16166489,0.98216084){\color[rgb]{0,0,0}\makebox(0,0)[lt]{\lineheight{1.25}\smash{\begin{tabular}[t]{l}$\bfv_{pr}$\end{tabular}}}}%
    \put(0,0){\includegraphics[width=\unitlength,page=2]{crossSecVels.pdf}}%
    \put(0.71686165,0.24140743){\color[rgb]{0,0,0}\makebox(0,0)[lt]{\lineheight{1.25}\smash{\begin{tabular}[t]{l}$\beta$\end{tabular}}}}%
  \end{picture}%
\endgroup%

		\caption{Cross-section diagram with angle of incidence $\beta$.}
		\label{fig:squareCantBeamCrossSecIlus}
	\end{subfigure}
	\caption{Example 3: Diagram of a square cantilever beam and relative wind flow for generic cross section.}
	\label{fig:squareCantBeamData}
\end{figure}

The problem consists in a cantilever beam with a square cross-section (width $a = \SquareCantBeama$ m), clamped on $x=0$  m and length $L = \SquareCantBeamL$ m, as shown in  Figure~\ref{fig:squareCantBeamIlus}. The beam is formed by a material with density $\rho = \SquareCantBeamrho$  kg/m$^3$ and Young modulus $E = \SquareCantBeamE$~kPa.

In this example the wind velocity is also considered to be a uniform field, however, the direction of the velocity vectors varies from $\bfc_2$ at $t=0$ s to a combination of $\bfc_2$ and $\bfc_3$ at $t=t_c$. The mathematical expression of the vector field is given by:
\begin{equation}\label{eq:resSquareCylindricalCantBeamWindVel}
	\bfv_a(\bfx, t)=
	\left\{\begin{array}{lr}
		v_{m}\frac{t}{t_c}\left( \cos(\frac{t}{t_c}\alpha_{m})\bfc_2 -\sin(\frac{t}{t_c}\alpha_{m})\bfc_3 \right)  & t \in [0, t_c] \\
		v_{m}\left( \cos(\alpha_{m})\bfc_2  -\sin(\alpha_{m})\bfc_3 \right)		   & t \in (t_c,+\infty) \\
	\end{array}
\right.
\end{equation}

where $\alpha_{m}=40 ^\circ$, $v_{m} = 2.8$ m/s and: $t_c=0$ (the steady case), $t_c = 1$ s (the non-steady case 1) and $t_c = 20$ s (the non-steady case 2).

\begin{figure}[htb]
	\centering
	\resizebox{.5\textwidth}{!}{\input{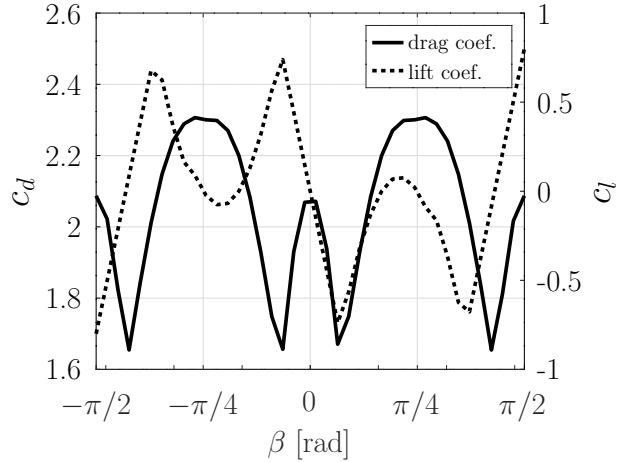}}	
 	\caption{Example 3: Functions considered for drag and lift coefficients.}
	\label{fig:squareCantBeamDragLift}
\end{figure}

In Figure~\ref{fig:squareCantBeamCrossSecIlus} a diagram of the deformed cross-section with the projected relative velocity is shown, where the angle of incidence $\beta$ is depicted. %
Realistic fluid-dynamic coefficients for a sharp-edge cross-section are considered based on \citep{Carassale2014}, for $Re=3.7 = \frac{v_m \cdot d_c}{\nu_f} \times 10^4$, with $d_c=a$. %
The plot of the drag and lift coefficient functions considered are shown in Figure~\ref{fig:squareCantBeamDragLift}. 

\subsubsection{Numerical results}

For the numerical resolution $\SquareCantBeamnumElem$ co-rotational frame elements with the consistent inertial formulation are considered. The Newmark method is used with stopping criteria $tol_r = \SquareCantBeamtolF0^{-10}$ and $tol_u = \SquareCantBeamtolU0^{-15}$.

Before obtaining the results for all the analysis cases, a study for different number of integration Gauss points is done. The non-steady case 1 is considered, and the displacement solutions obtained using the consistent formulation are compared at $t=\SquareCantBeamtimeError$ s. The solution obtained for 10 integration points is considered as \textit{reference}, and the relative errors are computed using the measure presented in Equation~\eqref{eq:resReleativeErrorLDcantBeam}.

\begin{figure}[htb]
	\begin{subfigure}{0.5\textwidth}
		\centering
		\resizebox{.95\textwidth}{!}{\input{SquareCantGaussAnalysis.tex}}
		\caption{Results for different numbers of Gauss integration points using displacements at $t=\SquareCantBeamtimeError$ s.}
\label{fig:squareCantBeam3DDynamicGauss}
	\end{subfigure}
	\begin{subfigure}{.5\textwidth}
			\centering
		\def\svgwidth{0.96\textwidth}
\begingroup%
  \makeatletter%
  \providecommand\color[2][]{%
    \errmessage{(Inkscape) Color is used for the text in Inkscape, but the package 'color.sty' is not loaded}%
    \renewcommand\color[2][]{}%
  }%
  \providecommand\transparent[1]{%
    \errmessage{(Inkscape) Transparency is used (non-zero) for the text in Inkscape, but the package 'transparent.sty' is not loaded}%
    \renewcommand\transparent[1]{}%
  }%
  \providecommand\rotatebox[2]{#2}%
  \newcommand*\fsize{\dimexpr\f@size pt\relax}%
  \newcommand*\lineheight[1]{\fontsize{\fsize}{#1\fsize}\selectfont}%
  \ifx\svgwidth\undefined%
    \setlength{\unitlength}{743.31139587bp}%
    \ifx\svgscale\undefined%
      \relax%
    \else%
      \setlength{\unitlength}{\unitlength * \real{\svgscale}}%
    \fi%
  \else%
    \setlength{\unitlength}{\svgwidth}%
  \fi%
  \global\let\svgwidth\undefined%
  \global\let\svgscale\undefined%
  \makeatother%
  \begin{picture}(1,0.68208291)%
    \lineheight{1}%
    \setlength\tabcolsep{0pt}%
    \put(0,0){\includegraphics[width=\unitlength,page=1]{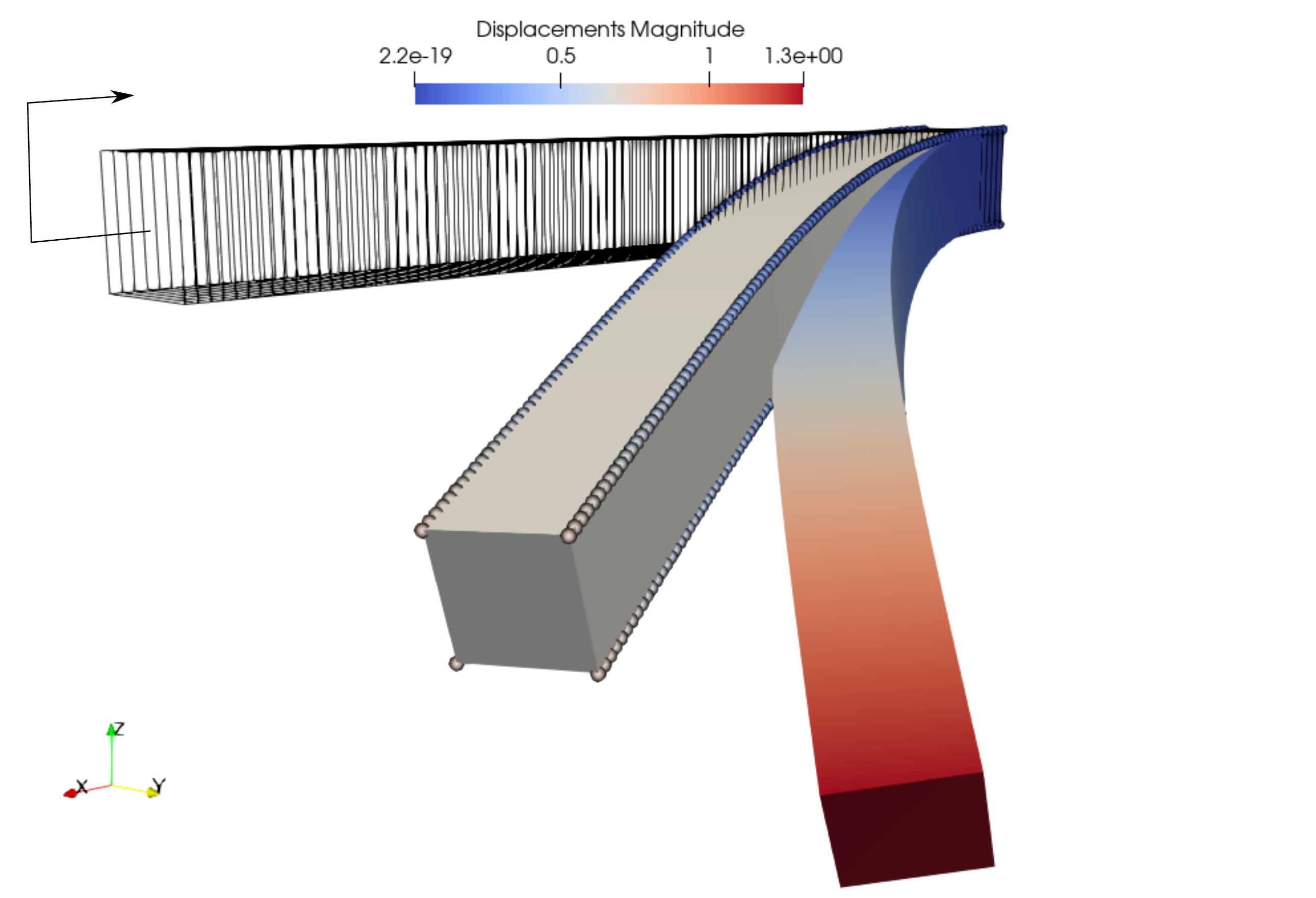}}%
    \put(0.10229507,0.60198468){\rotatebox{1.4471378}{\makebox(0,0)[lt]{\lineheight{1.25}\smash{\begin{tabular}[t]{l}($t=0$ s)\end{tabular}}}}}%
    \put(0.70105571,0.42816962){\rotatebox{-0.07587351}{\makebox(0,0)[lt]{\lineheight{1.25}\smash{\begin{tabular}[t]{l}case 1 ($t= 8$ s)\end{tabular}}}}}%
    \put(0.0095614,0.3962729){\rotatebox{-0.07587351}{\makebox(0,0)[lt]{\lineheight{1.25}\smash{\begin{tabular}[t]{l}case 1 ($t= 200$ s) \end{tabular}}}}}%
    \put(0,0){\includegraphics[width=\unitlength,page=2]{SquareCantDef.pdf}}%
    \put(0.22799695,0.07059318){\makebox(0,0)[lt]{\lineheight{1.25}\smash{\begin{tabular}[t]{l}steady case\end{tabular}}}}%
    \put(0,0){\includegraphics[width=\unitlength,page=3]{SquareCantDef.pdf}}%
  \end{picture}%
\endgroup%

		\caption{Steady case solution and non-steady case 1  deformed configurations at $t=\SquareCantBeamtimeError$ s and $t=200$ s.}
		\label{fig:squareCantBeam3DDef}
	\end{subfigure}
	\caption{Example 3: Gauss integration points analysis and plot of deformed configurations of the beam.}
\end{figure}

The results obtained are shown in Figure~\ref{fig:squareCantBeam3DDynamicGauss}. In Figure~\ref{fig:squareCantBeam3DDef} the deformed configurations of the beam for the steady case solution and the non-steady case 1 at times $t=\SquareCantBeamtimeError$ s and $t=200$ s are shown.  It is observed that a considerable curvature is present in the displacement field used, thus, a high nonlinearity is also present in the aerodynamic forces term. However, machine precision is reached using only six integration points.

\begin{figure}[htb]
	\begin{subfigure}{0.5\textwidth}
		\centering
		\resizebox{.95\textwidth}{!}{\input{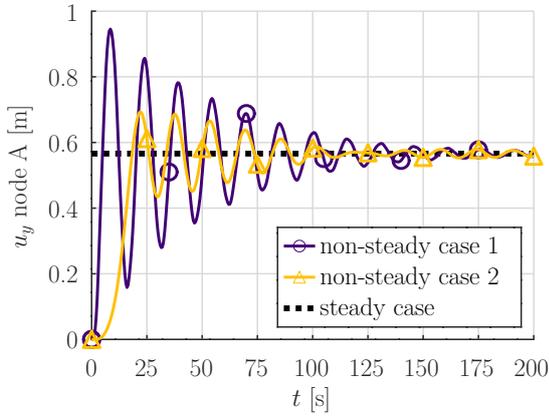}}		
		\caption{$u_y(t)$.}
		\label{fig:squareCantBeam3DDynamicUyA}
	\end{subfigure}
	\begin{subfigure}{.5\textwidth}
		\centering
		\resizebox{.95\textwidth}{!}{\input{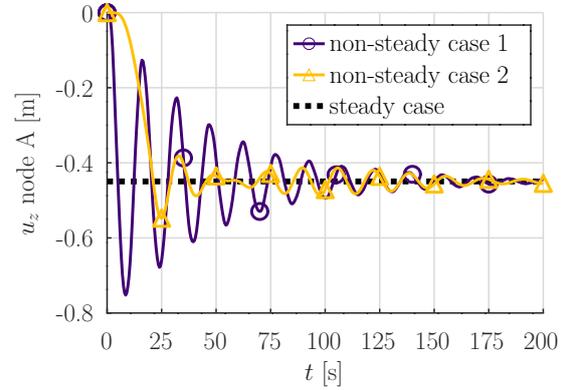}}		
		\caption{$u_z(t)$.}
		\label{fig:squareCantBeam3DDynamicUzA}
	\end{subfigure}
	\caption{Example 3: Displacements $u_y(t)$ and $u_z(t)$ of node A.}
	\label{fig:squareCantBeam3DDynamic}
\end{figure}

For the computation of the numerical solutions for all cases, four Gauss integration points are considered. The results obtained for displacements $u_z$ and $u_y$ of node A, are presented in Figures~\ref{fig:squareCantBeam3DDynamicUyA} and \ref{fig:squareCantBeam3DDynamicUzA}, respectively. It is observed that three dimensional geometrical nonlinearities are captured by the proposed formulation, while the non-steady solutions converge to the steady solution as expected. 

The results allow us to conclude that the proposed formulation can be used for problems with large rotations, and accurate results can be obtained, for the aerodynamic forces term, using four Gauss integration points.

\subsection{Example 4: simple propeller model}

\newcommand{\SimplePropellerL}{3}
\newcommand{\SimplePropellerd}{0.1}
\newcommand{\SimplePropellerEf}{2.1}
\newcommand{\SimplePropellerva}{1}
\newcommand{\SimplePropellerEr}{210}
\newcommand{\SimplePropellerrho}{6000}
\newcommand{\SimplePropellertolF}{1}
\newcommand{\SimplePropellertolU}{1}
\newcommand{\SimplePropellerdeltaT}{1}
\newcommand{\SimplePropellerfinalTime}{450}

In this example a simple propeller submitted to lift forces is considered. This problem is used to validate the dynamic response of the proposed formulation in a case with large displacements and rotations, comparing the numerical results with an analytic solution.

\subsubsection{Problem definition}

\begin{figure}[htb]
	\begin{subfigure}{0.5\textwidth}
		\def\svgwidth{1\textwidth}
		\centering
		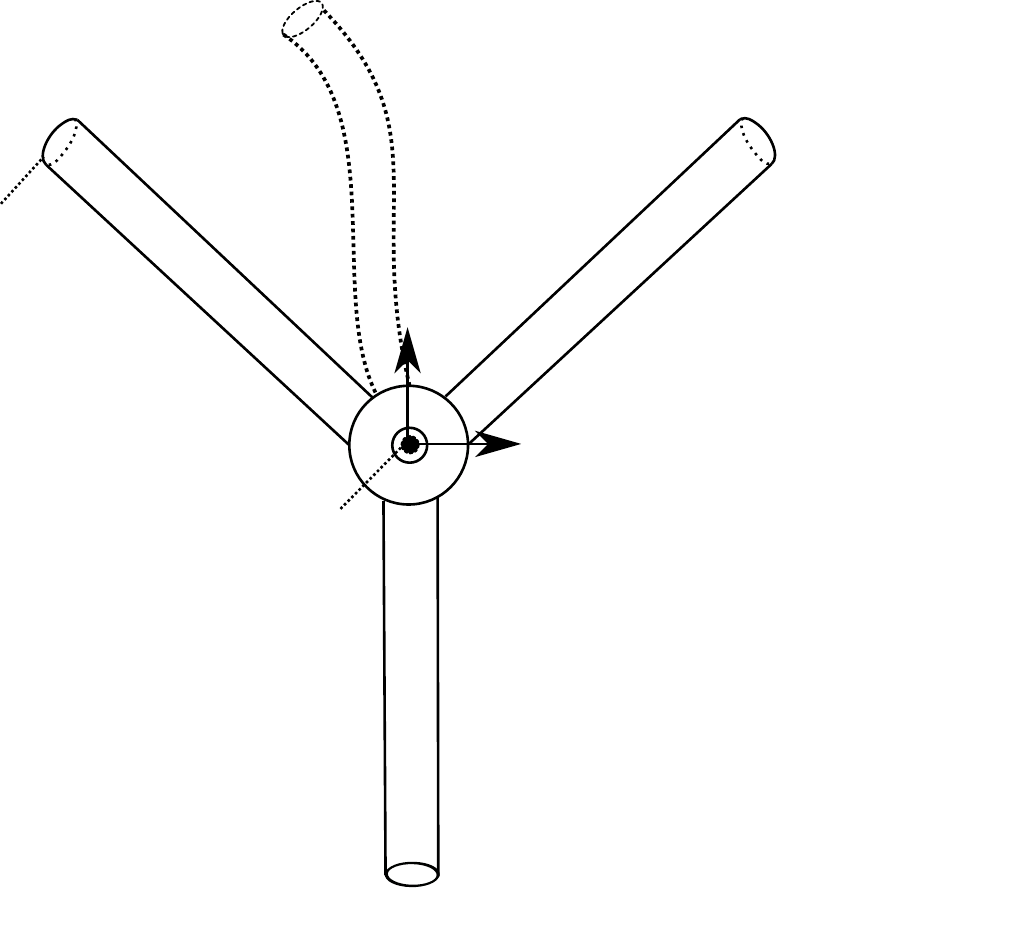
		\caption{Diagram of the geometry and rotations.}
		\label{fig:simplePropellerIlusGeometric}
	\end{subfigure}
	\begin{subfigure}{.5\textwidth}
		\def\svgwidth{0.9\textwidth}
		\centering
		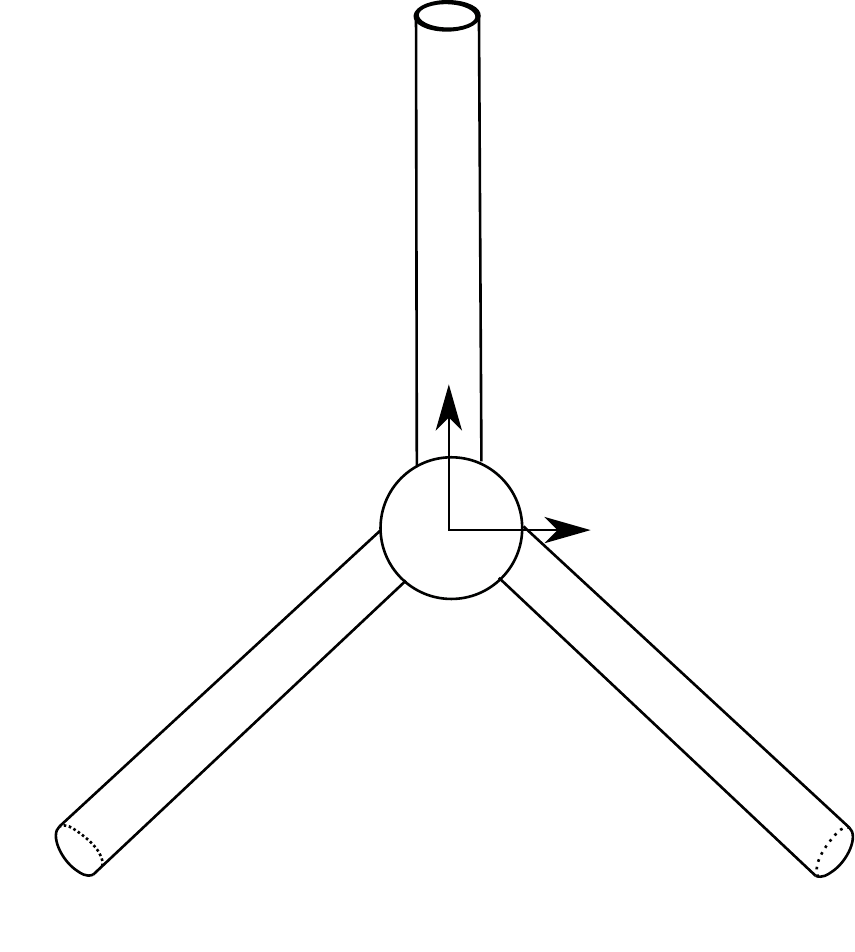
		\caption{Diagram of the fluid flow and lift forces.}
		\label{fig:simplePropellerIlusDynamic}
	\end{subfigure}
	\caption{Example 4: Diagram of a simple propeller.}
	\label{fig:simplePropellerIlus}
\end{figure}

The problem consists in a three-blade propeller, where each blade has a length $L = \SimplePropellerL $ m and a circular cross-section with diameter $d = \SimplePropellerd$ m, as shown in Figure~\ref{fig:simplePropellerIlusGeometric}. Regarding the stiffness of the blades, two cases are considered: a rigid case (with analytic solution) and a flexible case, providing considerably large rotations and bending of the blades. The density $\rho= \SimplePropellerrho $ kg/m$^3$ is considered. Regarding boundary conditions, the node O has five degrees of freedom fixed: the three displacements  and rotations $\theta_{y}$ and $\theta_{z}$. The rotation $\theta_{x,\text{O}}$ is free.

A uniform steady flow $\bfv_{a}=\SimplePropellerva~\bfc_1$ m/s is applied with synthetic aerodynamic coefficients $c_l = 0.2$ and $c_d = c_m = 0$. Given this, a uniform lift distributed force $\bff_l$ contained in the plane $\bfc_2$-$\bfc_3$ is induced, as shown in Figure~\ref{fig:simplePropellerIlusDynamic}.  These specific settings allow us to obtain an analytic solution for the rigid case.

\subsubsection{Numerical results: rigid case} 

For this case, and using the value selected for $\bfv_{a}$, it can be assumed that $\dot{\bfu} \ll \bfv_{a}$, therefore, $\bfv_{pr} \approx \bfv_{a}$. For the considered properties and boundary conditions, and for a Young modulus $E = \SimplePropellerEr$ GPa, the bending deformation of the blades can be neglected, allowing to obtain an analytic solution.

Considering $\theta_{x,\text{O}} \approx \theta_{x,\text{A}} = \theta_x$, the angular momentum balance equation:
\begin{equation}
\frac{1}{2}\rho_f c_l d ||\bfv_{a}||^2_2 \frac{L^2}{2} = \frac{1}{3}\rho L \pi \frac{d^2}{4}L^2 \ddot{\theta_x},
\end{equation}
and the homogeneous initial conditions, the analytic expression is obtained
\begin{equation}\label{eq:simplePropellerAnalytic}
 \theta_x(t) = \frac{3\rho_a c_l ||\bfv_{a}||^2_2}{2\rho L^3 \pi} t^2.
\end{equation}

For the numerical resolution the tolerances $tol_r = \SimplePropellertolF0^{-6}$ and $tol_u = \SimplePropellertolU0^{-12}$ are set. The $\alpha$-HHT numerical integration method with $\alpha = -0.05$ is used, with a time step increment set to $\Delta t = \SimplePropellerdeltaT$ s. Two and a half revolutions are modeled, with final simulation time $t_f=\SimplePropellerfinalTime$ s.

\begin{figure}[htb]
	\begin{subfigure}{0.5\textwidth}
		\centering
		\resizebox{\textwidth}{!}{\input{valPropRigidThetaXO.tex}}
		\caption{Results for $\theta_{x,\text{O}}(t)$.}
		\label{fig:simplePropellerRigidThetaX}
	\end{subfigure}
	\begin{subfigure}{.5\textwidth}
		\def\svgwidth{1\textwidth}
		\centering
\begingroup%
  \makeatletter%
  \providecommand\color[2][]{%
    \errmessage{(Inkscape) Color is used for the text in Inkscape, but the package 'color.sty' is not loaded}%
    \renewcommand\color[2][]{}%
  }%
  \providecommand\transparent[1]{%
    \errmessage{(Inkscape) Transparency is used (non-zero) for the text in Inkscape, but the package 'transparent.sty' is not loaded}%
    \renewcommand\transparent[1]{}%
  }%
  \providecommand\rotatebox[2]{#2}%
  \newcommand*\fsize{\dimexpr\f@size pt\relax}%
  \newcommand*\lineheight[1]{\fontsize{\fsize}{#1\fsize}\selectfont}%
  \ifx\svgwidth\undefined%
    \setlength{\unitlength}{665.25bp}%
    \ifx\svgscale\undefined%
      \relax%
    \else%
      \setlength{\unitlength}{\unitlength * \real{\svgscale}}%
    \fi%
  \else%
    \setlength{\unitlength}{\svgwidth}%
  \fi%
  \global\let\svgwidth\undefined%
  \global\let\svgscale\undefined%
  \makeatother%
  \begin{picture}(1,0.7621195)%
    \lineheight{1}%
    \setlength\tabcolsep{0pt}%
    \put(0,0){\includegraphics[width=\unitlength,page=1]{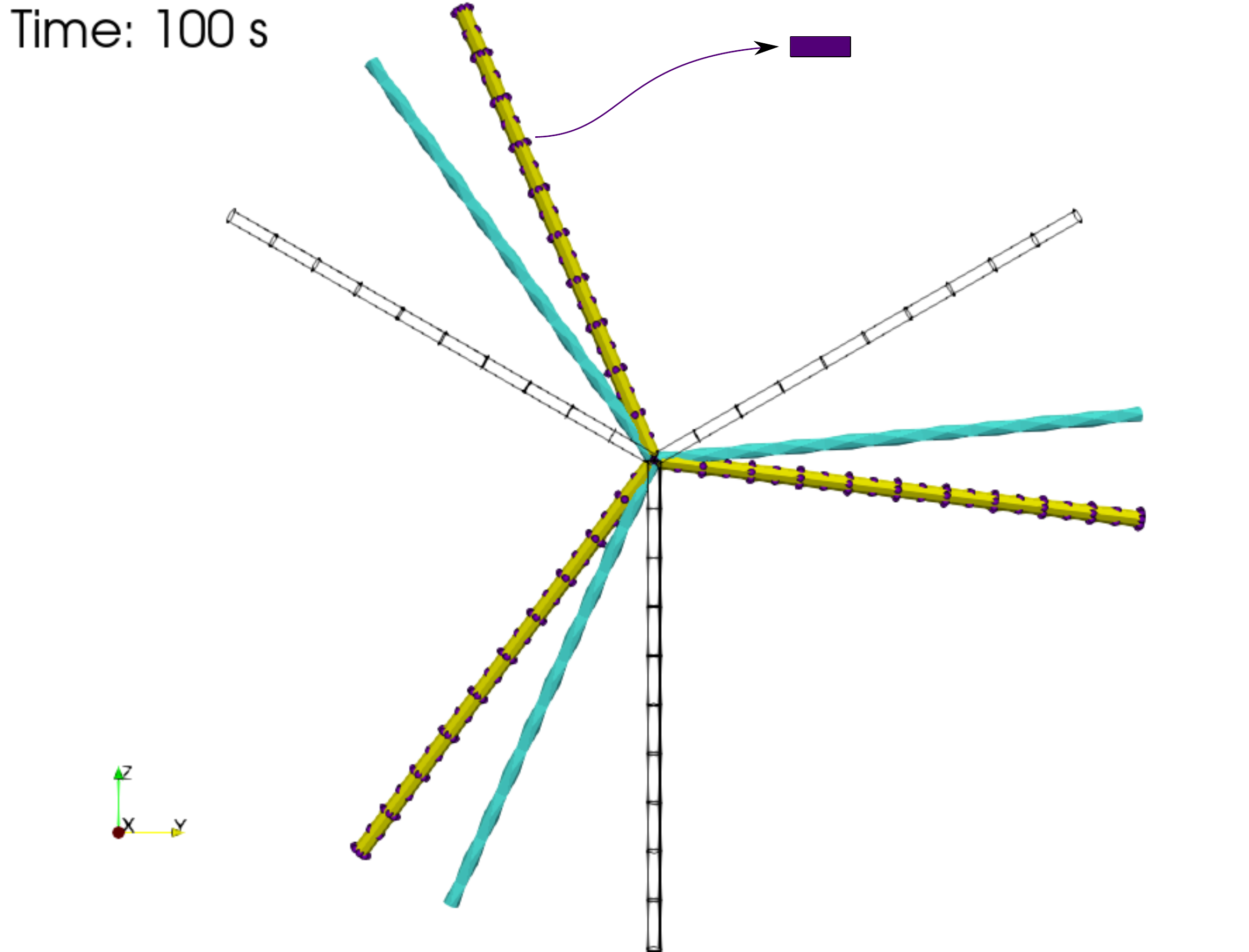}}%
    \put(0.69286944,0.71668312){\makebox(0,0)[lt]{\lineheight{1.25}\smash{\begin{tabular}[t]{l}consistent 1 elem.\end{tabular}}}}%
    \put(0,0){\includegraphics[width=\unitlength,page=2]{defPlotRigid.pdf}}%
    \put(0.73432835,0.62547936){\makebox(0,0)[lt]{\lineheight{1.25}\smash{\begin{tabular}[t]{l}consistent ref.\end{tabular}}}}%
    \put(0,0){\includegraphics[width=\unitlength,page=3]{defPlotRigid.pdf}}%
    \put(0.10681385,0.36049414){\makebox(0,0)[lt]{\lineheight{1.25}\smash{\begin{tabular}[t]{l}lumped. 1 elem\end{tabular}}}}%
    \put(0,0){\includegraphics[width=\unitlength,page=4]{defPlotRigid.pdf}}%
    \put(0.66585777,0.14536668){\makebox(0,0)[lt]{\lineheight{1.25}\smash{\begin{tabular}[t]{l}reference ($t=0$ s)\end{tabular}}}}%
  \end{picture}%
\endgroup%

		\caption{Deformed configurations at $t=100$ s.}
		\label{fig:simplePropellerRigidDef}
	\end{subfigure}
	\caption{Example 4: Rigid case results.}
	\label{fig:simplePropellerRigidResults}
\end{figure}

The results obtained for $\theta_{x,O}$ are shown in Figure~\ref{fig:simplePropellerRigidThetaX}, where it can be observed that the analytic solution matches those provided by the proposed consistent formulation, even using one element per blade. In contrast, the lumped formulation requires the use of five elements per blade to match the analytic solution. In Figure~\ref{fig:simplePropellerRigidDef} the deformed configurations are shown for time $t=100$ s.

\subsubsection{Numerical results: flexible case} 

The goal of this case is to test the proposed formulation for a highly-flexible propeller. To do so,  a Young modulus $E=\SimplePropellerEf$ kPa is considered. This local flexible behavior plays a key role in applications such as the design of morphing wings \citep{Zhang2020,Tsushima2019,DeBreuker2011a}, and represents a challenge for numerical methods.

\begin{figure}[htb]
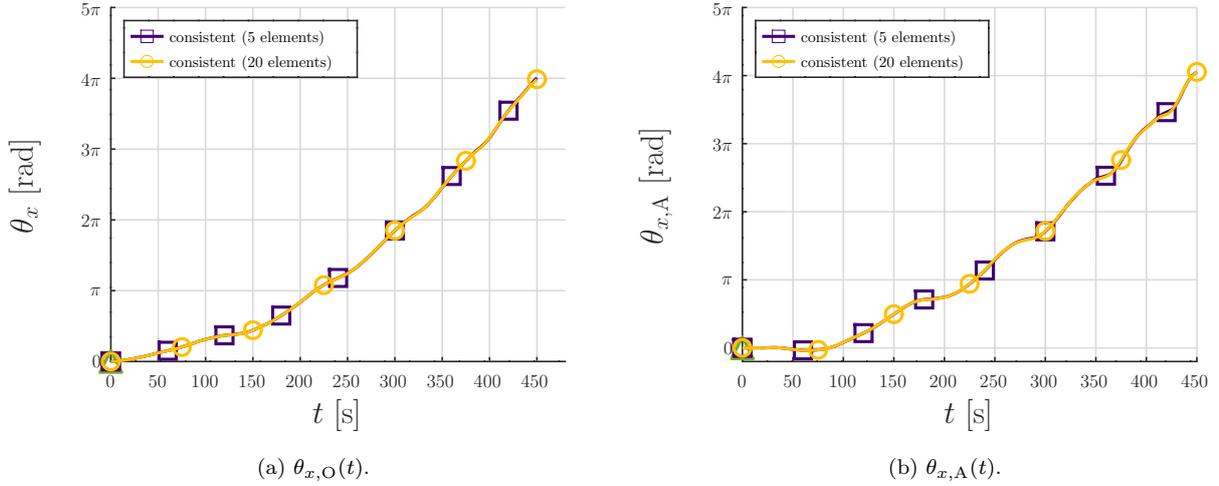

	\begin{subfigure}{0.5\textwidth}
		\centering
		\resizebox{\textwidth}{!}{\input{valPropFlexThetaXO.tex}}
		\caption{$\theta_{x,\text{O}}(t)$.}
		\label{fig:simplePropellerFlexThetaXO}
	\end{subfigure}
	\begin{subfigure}{.5\textwidth}
		\centering
		\resizebox{\textwidth}{!}{\input{valPropFlexThetaXA.tex}}
		\caption{$\theta_{x,\text{A}}(t)$.}
		\label{fig:simplePropellerFlexThetaXA}
	\end{subfigure}
	\caption{Example 4: Flexible case results of $\theta_{x,\text{A}}(t)$ and $\theta_{x,\text{O}}(t)$ rotations.}
	\label{fig:simplePropellerFlexThetas}
\end{figure}

\begin{figure}[htb]
	\centering
	\def\svgwidth{0.6\textwidth}
\begingroup%
  \makeatletter%
  \providecommand\color[2][]{%
    \errmessage{(Inkscape) Color is used for the text in Inkscape, but the package 'color.sty' is not loaded}%
    \renewcommand\color[2][]{}%
  }%
  \providecommand\transparent[1]{%
    \errmessage{(Inkscape) Transparency is used (non-zero) for the text in Inkscape, but the package 'transparent.sty' is not loaded}%
    \renewcommand\transparent[1]{}%
  }%
  \providecommand\rotatebox[2]{#2}%
  \newcommand*\fsize{\dimexpr\f@size pt\relax}%
  \newcommand*\lineheight[1]{\fontsize{\fsize}{#1\fsize}\selectfont}%
  \ifx\svgwidth\undefined%
    \setlength{\unitlength}{665.25bp}%
    \ifx\svgscale\undefined%
      \relax%
    \else%
      \setlength{\unitlength}{\unitlength * \real{\svgscale}}%
    \fi%
  \else%
    \setlength{\unitlength}{\svgwidth}%
  \fi%
  \global\let\svgwidth\undefined%
  \global\let\svgscale\undefined%
  \makeatother%
  \begin{picture}(1,0.7621195)%
    \lineheight{1}%
    \setlength\tabcolsep{0pt}%
    \put(0,0){\includegraphics[width=\unitlength,page=1]{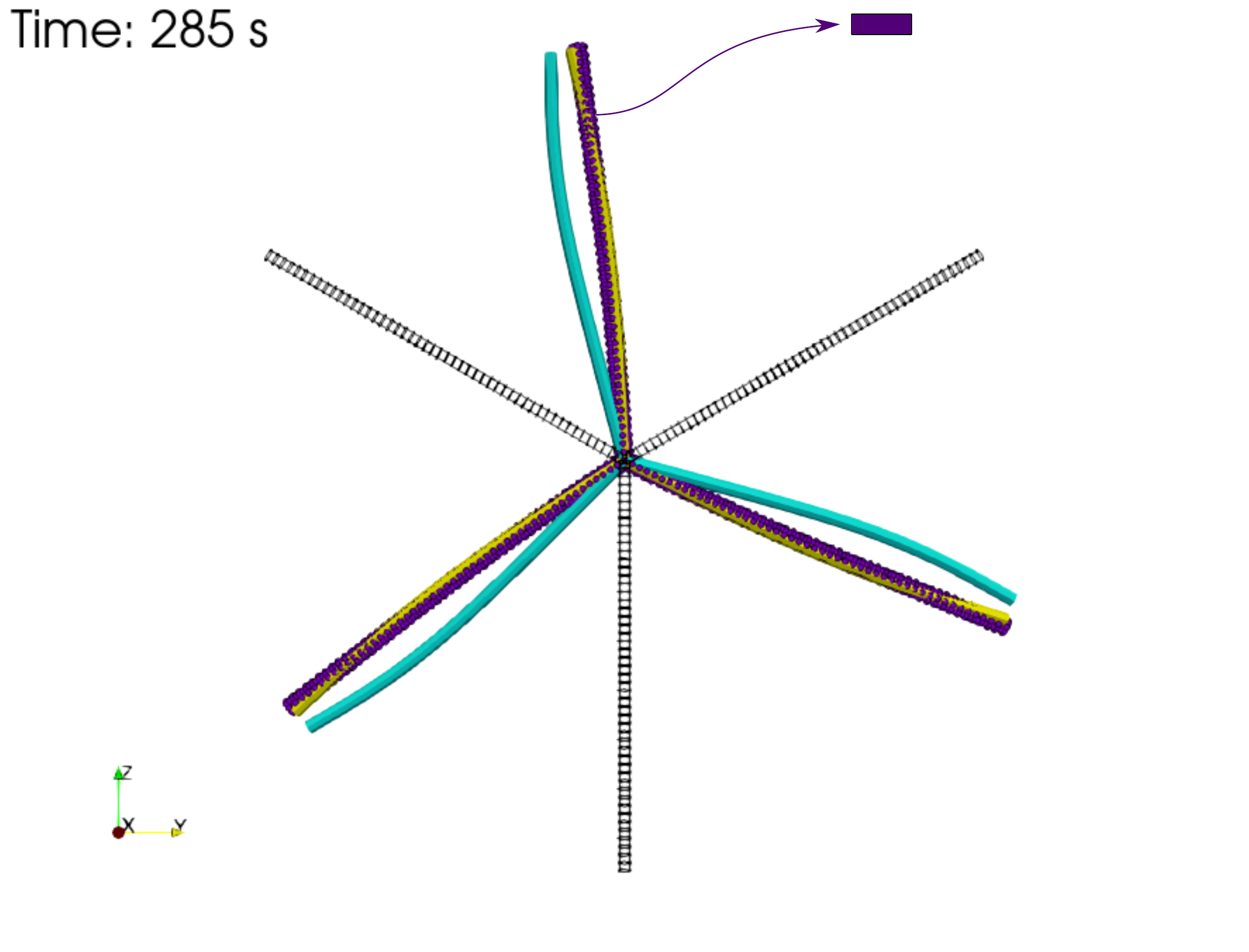}}%
    \put(0.74179246,0.73451232){\makebox(0,0)[lt]{\lineheight{1.25}\smash{\begin{tabular}[t]{l}consistent 5 elem.\end{tabular}}}}%
    \put(0,0){\includegraphics[width=\unitlength,page=2]{defFlexPlot.pdf}}%
    \put(0.77474036,0.6153533){\makebox(0,0)[lt]{\lineheight{1.25}\smash{\begin{tabular}[t]{l}consistent ref.\end{tabular}}}}%
    \put(0,0){\includegraphics[width=\unitlength,page=3]{defFlexPlot.pdf}}%
    \put(0.64394306,0.16139471){\makebox(0,0)[lt]{\lineheight{1.25}\smash{\begin{tabular}[t]{l}reference ($t=0$ s)\end{tabular}}}}%
    \put(0,0){\includegraphics[width=\unitlength,page=4]{defFlexPlot.pdf}}%
    \put(0.1101792,0.38355559){\makebox(0,0)[lt]{\lineheight{1.25}\smash{\begin{tabular}[t]{l}lumped. 5 elem\end{tabular}}}}%
  \end{picture}%
\endgroup%

	\caption{Example 4: Flexible case  deformed configurations at time $t=285$ s.}
	\label{fig:simplePropellerFlexDeformed}
\end{figure}

In this case, due to the bending deformation, the rotations of points O and A, shown in Figure~\ref{fig:simplePropellerIlusGeometric}, are considerably different. The numerical results obtained for the point O and point A using the consistent formulation are presented in Figures~\ref{fig:simplePropellerFlexThetaXO} and \ref{fig:simplePropellerFlexThetaXA}, respectively. Is reported that the lumped formulation requires 20 elements to match the reference solution for the first revolution, nevertheless a numerical divergence occurs at $t\approx 285$ s. This result indicates that for flexible elements with rotations larger than $2 \pi$ the use of the proposed inertial formulation is required to produce acceptable results. The deformed configurations obtained using consistent and lumped formulations at $t=285$ s are shown in Figure~\ref{fig:simplePropellerFlexDeformed}.

The results obtained let us conclude that the proposed consistent formulation provides accurate results for large rotations and considerable bending deformations.

\clearpage

\subsection{Example 5: S809 wind turbine }

\newcommand{\windTurbineL}{20}
\newcommand{\windTurbinedch}{1}
\newcommand{\windTurbineEeq}{14}
\newcommand{\windTurbineGeq}{35}
\newcommand{\windTurbinerhoeq}{1850}
\newcommand{\windTurbineva}{25}
\newcommand{\windTurbinetolF}{1}
\newcommand{\windTurbinetolU}{1}
\newcommand{\windTurbinedeltaT}{0.01}
\newcommand{\windTurbinefinalTime}{30}
\newcommand{\windTurbineelemBlade}{30}

\subsubsection{Problem definition}

\begin{figure}[htb]
	\begin{subfigure}{0.53\textwidth}
		\def\svgwidth{0.9\textwidth}
		\centering
		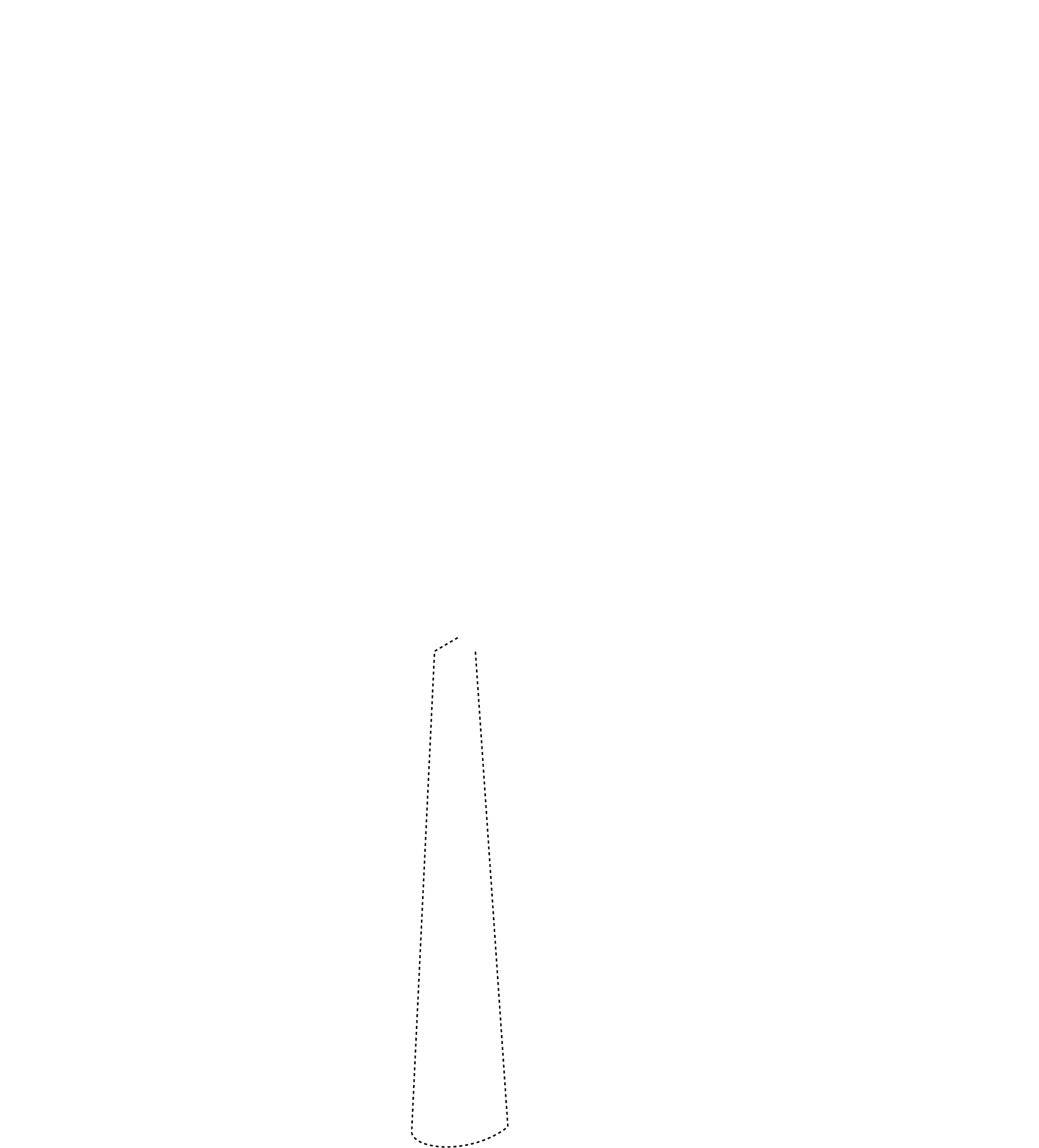
		\caption{Uniform wind turbine layout.}
		\label{fig:windTurbineIlus}
	\end{subfigure}
	\begin{subfigure}{.47\textwidth}
		\def\svgwidth{0.98\textwidth}
		\centering
		\resizebox{\textwidth}{!}{\input{S809crossSec.tex}}
		\caption{S809 airfoil shape extracted from \citep{Drela1989a}.}
		\label{fig:windTurbineS809crossSec}
	\end{subfigure}
	\caption{Example 5: Geometrical characteristics of uniform S809 wind turbine model. }
	\label{fig:windTurbineS809generalDim}
\end{figure}

In this example the capability of the proposed formulation to simulate the behavior of a realistic wind turbine is tested. The problem consists in an idealized wind turbine as shown in Figure~\ref{fig:windTurbineIlus}. Each blade has a uniform NERL S809 airfoil cross-section from \citep{Drela1989a}. The rotor has a radius $L_r=\windTurbineL$ m and the chord length $d_c= \windTurbinedch$ m remains constant from root to tip. A uniform constant wind velocity $\bfv_{a} = \windTurbineva~ \bfc_1$ m/s is considered.

For the material properties, considering \citep{FaccioJunior2019}  an equivalent Young modulus $E_{eq} = \windTurbineEeq$ GPa and an equivalent shear  modulus $G_{eq} = \windTurbineGeq$ GPa are considered. The density is $\rho_{eq} = \windTurbinerhoeq$ kg/m$^3$.

\subsubsection{Numerical results}
 
The numerical method employed is the $\alpha$-HHT with a time step $\Delta_t = \windTurbinedeltaT$ s and a final time $\windTurbinefinalTime$ s. The residual force and displacement tolerances are: $tol_r = \windTurbinetolF0^{-5}$ and $tol_u = \windTurbinetolU0^{-10}$. The spatial discretization of each blade is done using $\windTurbineelemBlade$ aerodynamic co-rotational elements.  
 
The aerodynamic coefficients for an angle of incidence between (-10$^\circ$, 20$^\circ$) are extracted from \citep{Drela1989a} and an extension of these coefficients for the range between -30$^\circ$ and 90$^\circ$ is done based on \citep{jonkman2003modeling}. 

\begin{figure}[htb]
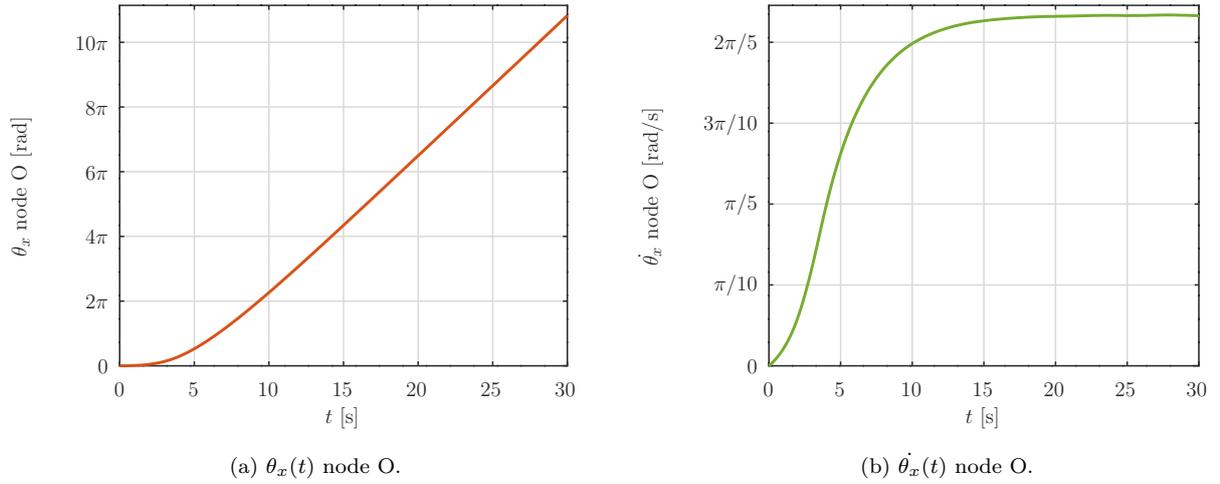

	\begin{subfigure}{0.5\textwidth}
		\centering
		\resizebox{\textwidth}{!}{\input{thetaX.tex}}
		\caption{$\theta_x(t)$ node O.}
		\label{fig:windTurbineThetaXO}
	\end{subfigure}
	\begin{subfigure}{.5\textwidth}
		\centering
		\resizebox{\textwidth}{!}{\input{thetadotX.tex}}
		\caption{$\dot{\theta_x}(t)$ node O.}
		\label{fig:windTurbineThetadotXO}
	\end{subfigure}
	\caption{Example 5: Azimuths $\theta_x$ angle and velocity rotation.}
	\label{fig:windTurbineThetaX0}
\end{figure}

The numerical results presented in Figure~\ref{fig:windTurbineThetaX0} provide an adequate dynamic response. The azimuths angle  $\theta_x$ of node O shown in Figure~\ref{fig:windTurbineThetaXO} increases until drag forces balances the lift momentum, and the angular acceleration tends to zero as shown in Figure~\ref{fig:windTurbineThetadotXO}.  

\begin{figure}[htb] 
		\centering
		\resizebox{.5\textwidth}{!}{\input{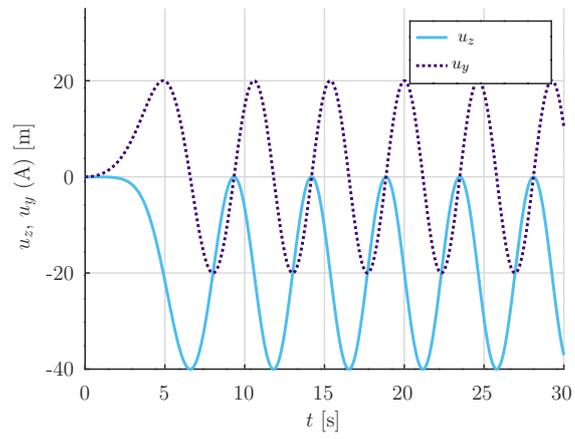}}
		\caption{Example 5: Displacements $u_y(t)$ and $u_z(t)$ of node A.}
		\label{fig:windTurbineDispsA}
\end{figure}

The horizontal $u_y$ and vertical $u_z$ displacements of node A are plotted in Figure~\ref{fig:windTurbineDispsA}. It can be noted that harmonic oscillation trajectory is presented in accordance with results depicted in Figure~\ref{fig:windTurbineThetaX0}. In addition more than 5 revolutions are modeled with the proposed formulation, allowing us to conclude that the formulation can be used for problems with realistic properties.

\clearpage

\section{Conclusions}\label{sec:conclusions}

In this article a new formulation for the numerical analysis of frame structures submitted to aerodynamic forces is presented. The methodology is based on the co-rotational approach for computing the aerodynamic forces in the deformed configuration for large displacements and rotations. These co-rotational aerodynamic forces are integrated with co-rotational inertial and internal forces, using the principle of virtual work, providing a set of nonlinear governing equations. The proposed formulation and a numerical resolution procedures were implemented in the open-source FEM library ONSAS. The formulation and its implementation are validated through the resolution of five examples.

The proposed formulation represents a simple yet accurate tool for the simulation of fluid-structure interaction problems, without the requirement of performing complex and computationally demanding analyses. The open-source implementation of the formulation, and its resolution method, represent an attractive tool for early stages of frame structures design, for which the effect of fluid flows is relevant.

In Example 1 a simple cantilever beam problem with semi-analytic solution was considered. This problem was used to validate the formulation and the implementation for small displacements and steady and non-steady flows. The results obtained let us conclude that the proposed formulation provides accurate results.

In Example 2 a flexible cantilever beam submitted to a fluid flow with an oscillating velocity field was considered. This problem is used to validate the formulation for large displacements of the frame elements. By comparing the lumped and consistent mass approaches we conclude that the proposed formulation allows to simulate movements with large displacements using a reduced number of elements.

In Example 3 a cantilever beam with a square cross-section submitted to a fluid flow with uniform velocity and rotating direction was considered. For the drag and lift coefficients realistic functions were proposed using reference literature. Additionally, an for different Gauss integration points is also presented. The results let us conclude that the formulation is able to provide accurate results for the non-steady cases, for large displacements and considerable three-dimensional rotations. Moreover, for this case, the use of four Gauss integration points provides accurate results, while six or more points provides results with machine precision in the computation of the aerodynamic forces.

In Example 4 a three-blade propeller submitted to a uniform wind flow producing large rotations was considered. Two cases were defined for the stiffness of the blades (rigid and flexible), where the analytic solution is presented for the rigid case. The results obtained let us conclude that the proposed formulation provides accurate results and requires a reduced number of elements. For the flexible case, the lumped approach is not able to solve the problem, while the proposed consistent approach is able to provide adequate results.

Finally, the results obtained in Example 5 let us conclude that the proposed formulation can be applied to realistic problems. In this example a realistic wind turbine with NERL S809 blades is considered, with drag, lift and moment coefficient functions based on reference literature. The results obtained let us conclude that the proposed formulation might be used to solve real Engineering problems under the hypotheses considered in the example, at early stages of structural design.

Several research directions remain open, including: the extension of the present formulation to consider eccentric aerodynamic and mass centers; test the proposed formulation against results provided by other more computationally demanding FSI approaches; or extend the model of aerodynamic forces to reproduce more complex phenomena, such as flow-induced vibrations.

\section*{Acknowledgments}

The authors would like to thank \textit{Comisión Sectorial de Investigación Científica} of \textit{Universidad de la República} and \textit{Comisión Académica de Posgrado} for their financial support and \textit{Agencia Nacional de Investigación e Innovación} for the financial support through project {FSE\_1\_2016\_1\_131837}. The authors also want to thank Professors M. Forets, G. Usera and J. B. Bazzano, from Universidad de la República, for their valuable contributions during different stages of this work. Finally, the authors thank the ONSAS developers and contributors, specially, Prof. Jean-Marc Battini for his contributions to  the co-rotational internal forces function.

\bibliography{paper.bib}

\end{document}